\documentclass[aip,reprint,floatfix]{revtex4-1}
\usepackage{lineno,hyperref}
\usepackage{multirow}
\usepackage{xcolor}
\usepackage{graphicx}
\usepackage{dcolumn}
\usepackage{bm}
\usepackage[utf8]{inputenc}
\usepackage[T1]{fontenc}
\usepackage{mathptmx}
\usepackage{amssymb}
\usepackage{url}

\draft 

\begin{document}


\title[Chaos 31, 033130 (2021); doi: 10.1063/5.0023427]{Sequential seeding in multilayer networks} 



\author{Piotr Bródka}
\email{piotr.brodka@pwr.edu.pl}%
\homepage{http://piotrbrodka.pl}
\affiliation{Wroc\l{}aw University of Science and Technology, Department of Computational Intelligence, Wybrze\.ze Wyspia\'nskiego 27, Wroc\l{}aw, 50-370, Poland}
\affiliation{West Pomeranian University of Technology, Faculty of Computer Science and Information Technology, Zolnierska 49, Szczecin, 71-210, Poland}

\author{Jarosław Jankowski}%
\affiliation{West Pomeranian University of Technology, Faculty of Computer Science and Information Technology, Zolnierska 49, Szczecin, 71-210, Poland}

\author{Radosław Michalski}
\affiliation{Wroc\l{}aw University of Science and Technology, Department of Computational Intelligence, Wybrze\.ze Wyspia\'nskiego 27, Wroc\l{}aw, 50-370, Poland}%
\affiliation{West Pomeranian University of Technology, Faculty of Computer Science and Information Technology, Zolnierska 49, Szczecin, 71-210, Poland}


\date{\today}

\begin{abstract}
Multilayer networks are the underlying structures of multiple real-world systems where we have more than one type of interaction/relation between nodes: social, biological, computer, or communication, to name only a few. In many cases, they are helpful in modelling processes that happen on top of them, which leads to gaining more knowledge about these phenomena. One example of such a process is the spread of influence. Here, the members of a social system spread the influence across the network by contacting each other, sharing opinions or ideas, or - explicitly - by persuasion. Due to the importance of this process, researchers investigate which members of a social network should be chosen as initiators of influence spread to maximise the effect. In this work, we follow this direction, develop and evaluate the sequential seeding technique for multilayer networks. Until now, such techniques were evaluated only using simple one layer networks. The results show that sequential seeding in multilayer networks outperforms the traditional approach by increasing the coverage and allowing to save the seeding budget. However, it also extends the duration of the spreading process.
\end{abstract}


\maketitle 

\begin{quotation}

Sequential seeding is a node activation strategy for influence spreading, which distributes activations over time instead of performing all of them at once. It proved its superiority in a majority of seeding scenarios. However, the research until now was limited to simple one layer static networks reflecting only one type of relation. 

In this paper, we address that gap by extending sequential seeding to a multilayer network scenario. We have performed an extensive evaluation using four real networks and six synthetic (three random networks and three multilayer networks) of various sizes and with a various number of layers.

Our results show that sequential seeding outperforms the traditional approach by increasing the coverage and increasing the duration of spread, confirming findings from previous research for one layer networks. What is more, we have evaluated additional aspects of sequential seeding, namely, the savings in the seeding budget. This aspect has not been evaluated in previous research.

The findings presented in this work allow redesigning influence spreading strategies to increase the coverage with limited seeding budget, allowing for more effective spreading in multilayer networks.
\end{quotation}

\section{Introduction}
The influence maximisation problem challenges the researchers for more than fifteen years~\cite{kempe2003maximizing}. In its basic formulation, one needs to find a set of nodes in a complex network that activates the maximum number of nodes for a given influence spreading model. These nodes, usually called a seed set, are activated at the beginning of the process, and throughout the iterations, they spread the influence across the network. Unfortunately, the discovery of the best seed set equals a complete and in-depth evaluation of the spreading capabilities of nodes, which is an extremely hard and time-consuming task. Thus, multiple heuristics have been proposed which allow to find maybe not the best but good enough seed set. In a simplistic scenario, two primary factors contribute to the problem: the network topology and the influence model. However, when considering more realistic situations, additional factors should be taken into account, such as varying cost of acquiring/activating seed nodes or more complex network structures like temporal or multilayer networks.

Apart from that, recent research demonstrated that for static networks, the concurrent seed nodes activation is superseded by sequential activation of nodes from the seed set. The roots of this approach, called sequential seeding~\cite{jankowski2017balancing} (Section~\ref{sec:sq}), come from decision making, where information about the consequences of prior decisions should be gathered before making the next one. In sequential seeding, that information relates to the observation of how influence cascades spread in the network, and before selecting seed nodes for next activations, the knowledge on the current state of the process is taken into account. This method contrasts single stage seeding (Section~\ref{sec:ss}) with respect to how the activations are distributed, but what is worth underlining, sequential seeding is actually a meta-method, since it relates to the way how to activate the nodes independently of the actual seed set construction method. Still, the ordering of nodes to be activated can be generated by any heuristic (Section~\ref{sec:seed_selection}), depending on the time required to generate the seed set. For the independent cascade model (Section~\ref{sec:ICM}) and static networks, sequential seeding demonstrated its superiority over single stage activation\cite{jankowski2018probing}.

In this work, our goal is to extend the knowledge about the performance of the sequential seeding method by adding another degree of complexity: we are investigating sequential seeding in multilayer networks~\cite{kivela2014multilayer}. Multilayer networks are often a typical abstraction of the way how we interact with each other or how other complex systems work. For instance, in the case of human interactions, each layer in these networks can represent different means of communication like face meetings, phone calls, text messages, e-mails, or WhatsApp communication. Each of these layers can have different properties, structure, and importance. For instance, face to face meetings are crucial for disease spreading, while phone calls or online presence nowadays are more important in information and influence spreading. This also implies how influence spreads and albeit some works already looked at how different combinations of layers impact the spread~\cite{michalski2013convince,boccaletti2014structure,de2016physics,brodka2020interacting} and some works already proposed several heuristics for influence maximisation in this setting~\cite{basaras2017identifying,huang2020identifying,erlandsson2017seed}, sequential seeding in a multilayer scenario has not been studied so far. This is why this work attempts to fill this gap by investigating this approach by using multiple real world and synthetic networks.

The structure of this manuscript is as follows. In the next section, we present the methodological background. In Section\ref{sec:experiment} the experimental space is defined. Section~\ref{sec:results} demonstrates and discusses the results, whilst the last section concludes our findings.

\section{Methodology}
In this section, we will briefly introduce all methods and techniques we have used in our research. 

\subsection{Multilayer network}
\label{sec:multilayer}
\begin{figure}
 \centering
 \includegraphics[width=.4\textwidth]{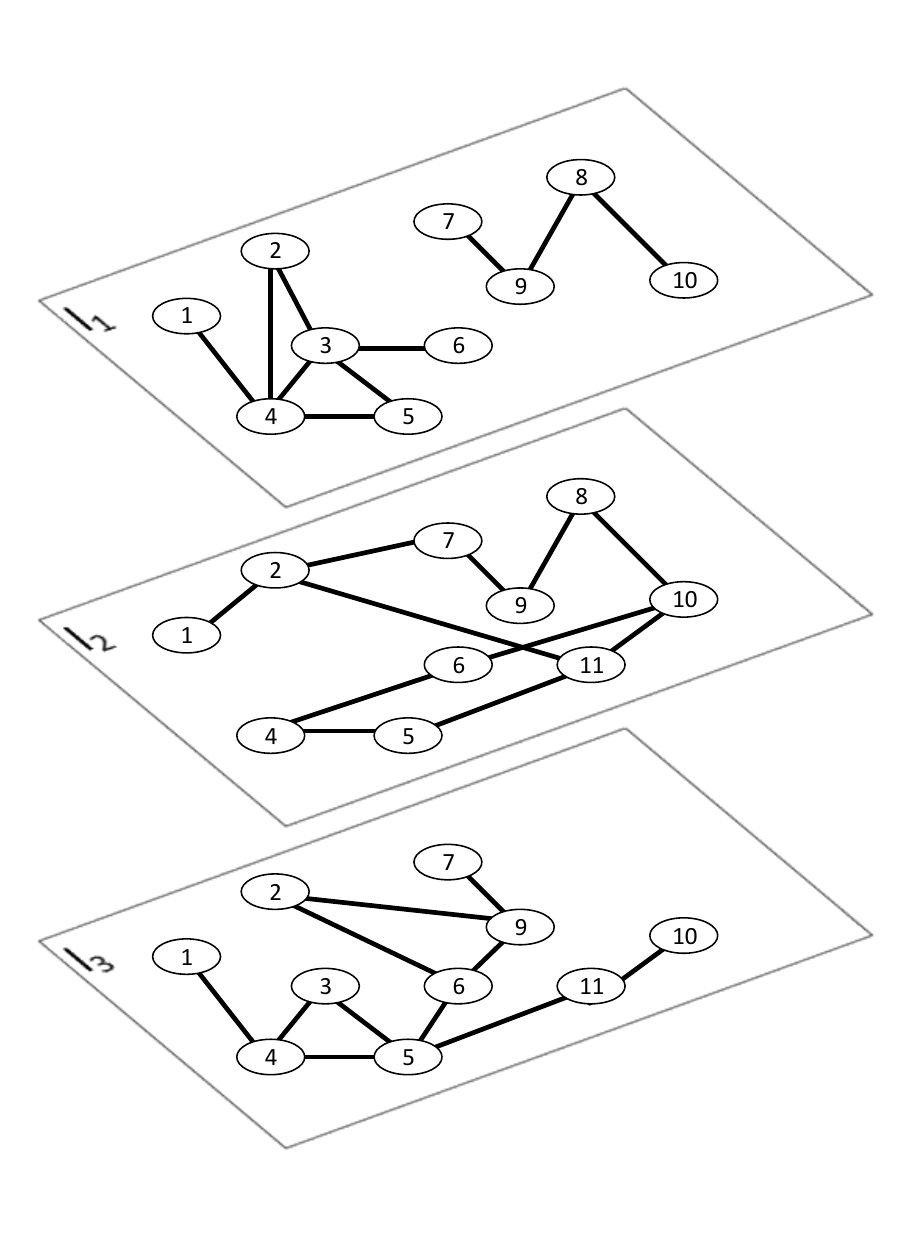}
 \caption{Toy example of a multilayer network}
 \label{fig:network}
\end{figure}

In the literature, many different definitions of multilayer networks exist \cite{magnani2011ml, Brodka2018, brodka2020interacting}; however, in this work, we use the definition of the multilayer network similar to proposed by Kivela at al.\cite{kivela2014multilayer}. 

Multilayer network is defined as quadruple $M = (N,L,V,E)$, where 
\begin{itemize}
 \item $N$ is a set of actors,
 \item $L$ is a set of layers, 
 \item $V$ and a set of nodes, $V \subseteq N \times L$,
 \item $E$ is a set of edges $(v_1, v_2): v_1, v_2 \in V$, and if $v_1=(n_1, l_1)$ and $v_2=(n_2, l_2) \in E$ then $l_1=l_2$.
\end{itemize}

An example of a multilayer network is presented in Fig.~\ref{fig:network}, where $L = \{l_1, l_2, l_3\}$, $N = \{1, 2, \dots, 11\}$, and $((1,l_2),(2,l_2))$ is an example of an edge in $E$. 

\subsection{Independent Cascade model}
\label{sec:ICM}
To simulate influence spreading in a network, we have used Independent Cascade model (ICM)~\cite{goldenberg2001talk}. In this model, each activated node has one chance to activate its neighbours with a defined propagation probability (PP). If a node activates its neighbours, they will have the chance to do the same in the next iteration of the process.
Since the initial version of this model has been proposed for simple one layer networks, we had to adjust it to a multilayer scenario, similarly to \cite{erlandsson2017seed}.

In a multilayer network, each newly activated actor will attempt to activate all its neighbours on each layer independently. This reflects the situation where, for example, we work (first layer) and play football (second layer) with someone. As a consequence we have more chances to influence that person due to multiple channels of interaction.

Additionally, if the actor is activated on one layer, it becomes active on all of them, i.e., it does not matter if someone convinces us at work or during football practice, we will have the same opinion everywhere.
Apart from these two changes, the multilayer ICM works the same as the original one layer model.

\subsection{Seed selection strategy}
\label{sec:seed_selection}

\begin{table}
\centering
\begin{tabular}{p{3.5cm}|p{.3cm}p{.3cm}p{.3cm}p{.3cm}p{.3cm}p{.3cm}p{.3cm}p{.3cm}p{.3cm}p{.4cm}p{.3cm}}
 Actor & 1 & 2 & 3 & 4 & 5 & 6 & 7 & 8 & 9 & 10 & 11 \\ \hline
 Degree Centrality & 3 & 7 &6 & \textcolor{blue}{\textbf{9}} & \textcolor{cyan}{\textbf{8}} & 6 &4 &4 & 7 &5 &5\\
 Neighbourhood Size & 2 & \textcolor{blue}{\textbf{7}} & 4 & 5 & 4 & \textcolor{cyan}{\textbf{6}} & 2 & 2 & 4 & 3 & 3 \\ 
\end{tabular}
\caption{The values of degree centrality and neighbourhood size measures for each actor for exemplary multilayer network (Fig. \ref{fig:network}) }
\label{tab:toyNetMeasures}
\end{table}

As mentioned in the previous section, there are a number of seed selection strategies, and since evaluating all of them is not the aim of our research, we have decided to select three simple and most commonly used seed selection strategies to observe if they have any impact on our results. 
 
The first one is a multilayer\textit{degree centrality}, i.e. the number of edges adjacent to each actor on all layers. The second one is the multilayer \textit{neighbourhood size}, i.e. the number of all distinct actors each actor is linked to in all layers \cite{berlingerio2011foundations, magnani2011ml}. Both measures may seem to be similar, since for \textit{neighbourhood size} we are counting neighbours and for \textit{degree centrality} we are counting edges connecting to those neighbours (in case a node has only one edge to each neighbour, both measures would have the same value). However, they both can yield very different results in terms of actors importance ranking. For example, for our toy network presented in Fig.~\ref{fig:network} we can see that the most important actors according to degree centrality are actors $4$ and $5$, while if we use the neighbourhood size, actors $7$ and $6$ are the most important (Table \ref{tab:toyNetMeasures}).
The third seed selection strategy was \textit{random} seed selection. 

Using each strategy for each network, we have created the ranking of actors with the most important on the top and the least important on the bottom. If two actors had the same value of measure, they were ordered according to actor id, e.g., for our toy example, actors $3$ and $6$ have the same value of degree centrality, but in the final ranking actor $3$ would be higher than actor $6$.
Next, the rankings were saved, and during simulations, the same rankings have been used regardless of other simulation parameters. This was especially important in the case of random seed selection.

\subsection{Single stage seeding}
\label{sec:ss}

\begin{figure*}
 
 \centering
 \includegraphics[height=.8\textheight]{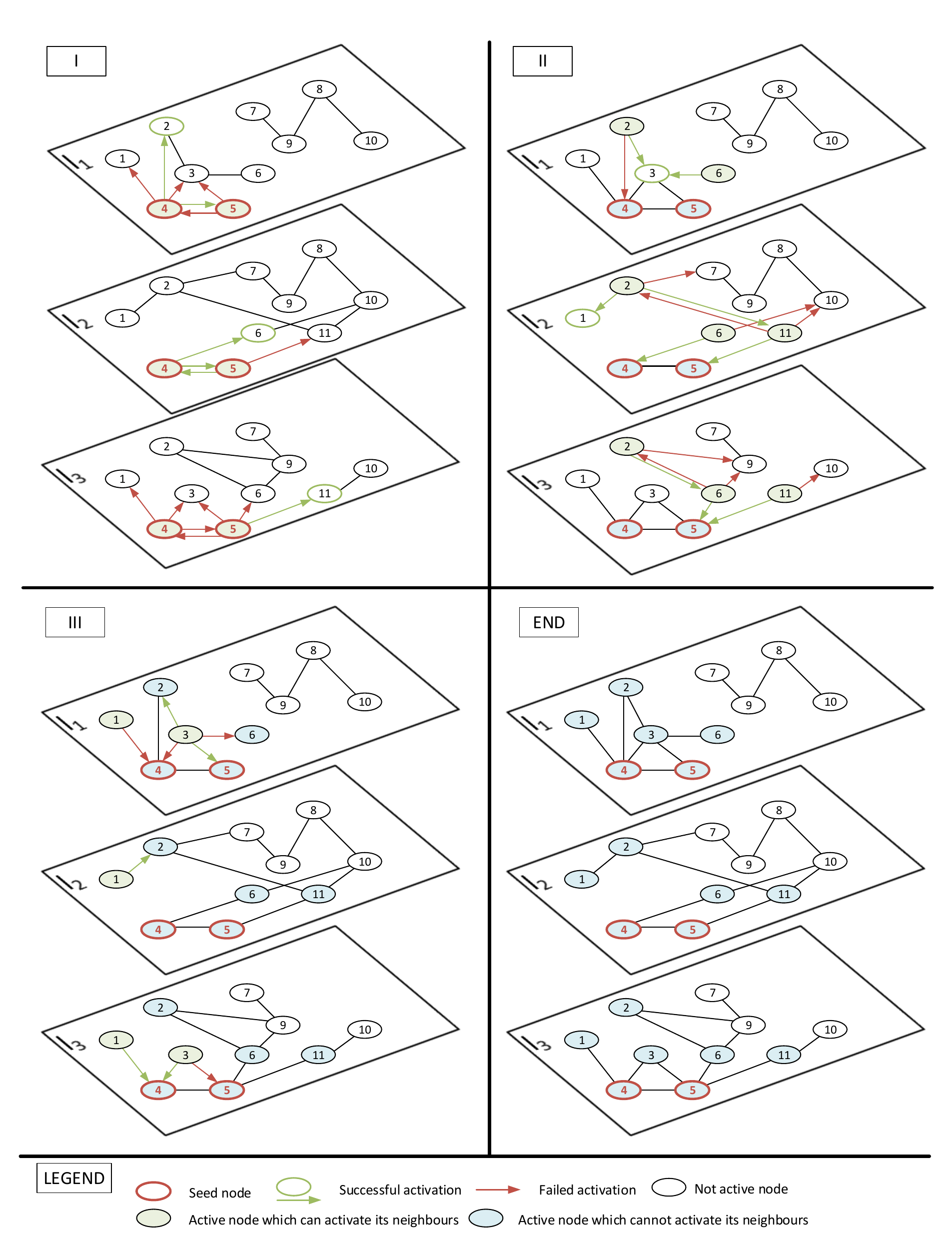}
 \caption{A toy example of single stage seeding in multilayer network (toy network from Fig.~\ref{fig:network}). Before we start spreading we calculate degree centrality, rank actors according to this measure and select two actors with the highest degree centrality, i.e., actor 4 and 5, as seeds (see Table~\ref{tab:toyNetMeasures}). Next we activate both of them and start spreading process simulated using Independent Cascade model. It finishes after three iterations with 7 actors activated (63.6\% of network).}
 \label{fig:ss_example}
\end{figure*}

Single stage seeding (SS) is a traditional approach to spreading initiation in one layer and multilayer networks \cite{salehi2015spreading, erlandsson2017seed}. 

In this technique, before we initiate the process, we create a seed set consisting of one or more actors (or simply nodes in one layer network) selected using some heuristic, e.g. actors with the highest degree centrality\cite{musial2009user}, k-shell centrality\cite{hu2013new}, page rank\cite{page1999pagerank}, neighbourhood size, VoteRank\cite{zhang2016identifying} or by simply selecting them at random.

When having a seed set defined in a single stage seeding approach, we activate all actors in the set at once (in a single stage) and allow them to influence other actors in the network. At this point, we do not have any additional control over the spreading and simply wait until it ends, i.e., there are no more actors that can be activated. The toy example of this process is presented in Fig.~\ref{fig:ss_example}.

\subsection{Sequential seeding}
\label{sec:sq}

\begin{figure*}
 \centering
 \includegraphics[width=\textwidth]{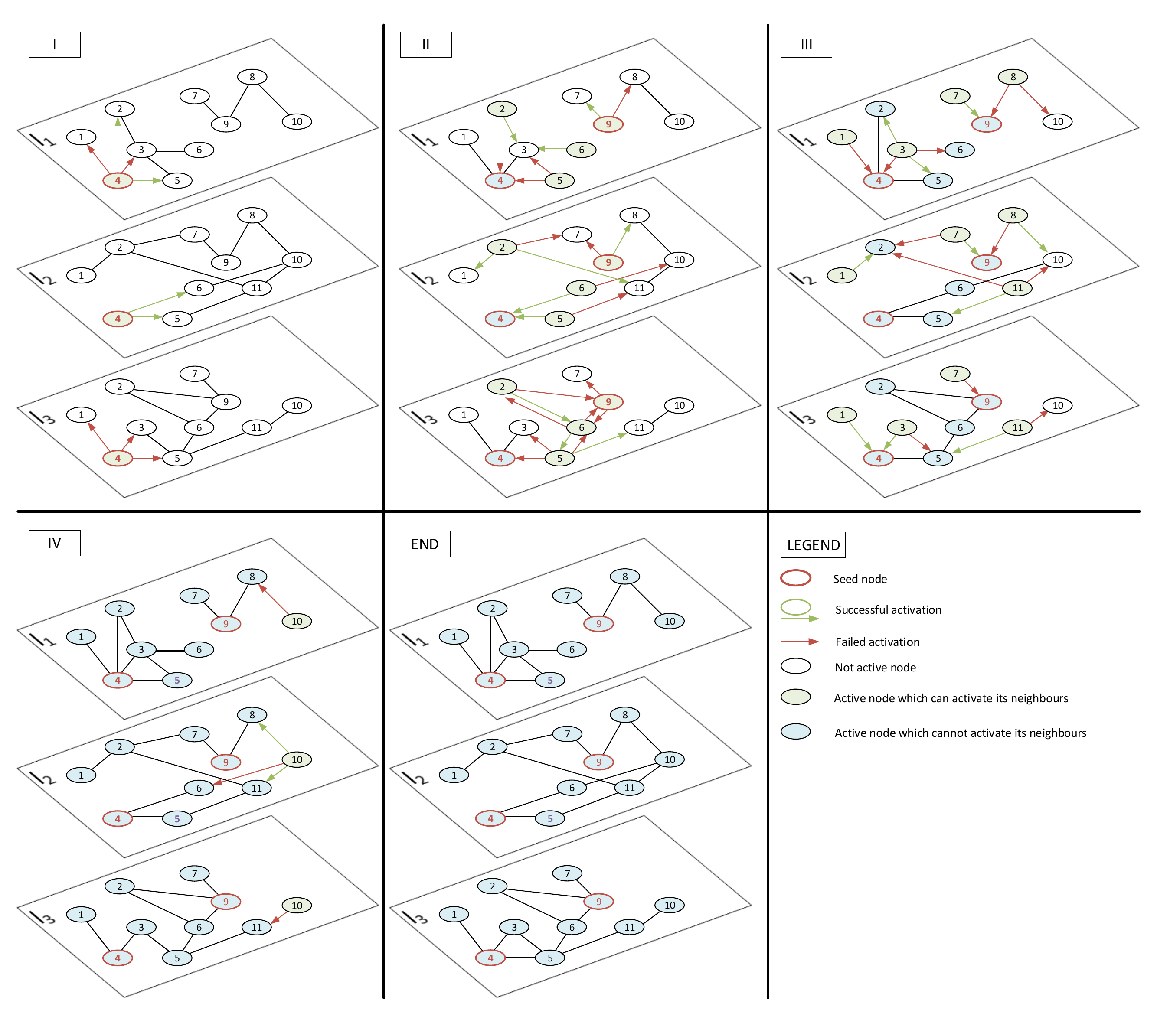}
 \caption{A toy example of sequential seeding in multilayer network (toy network from Fig. \ref{fig:network}). Before we start spreading we calculate degree centrality, rank actors according to this measure and select actor with the highest degree centrality (Table \ref{tab:toyNetMeasures}), i.e., actor 4 as the first seed. Next, we activate actor 4 and start ICM spreading process. The pattern of activations is exactly the same as in single stage seeding example (Fig. \ref{fig:ss_example}). However, due to the fact that we have used only one seed (half of our seeding budget) to start the spreading we still can add one more seed in the second stage (second iteration). We select the first non active actor on degree ranking (i.e. the actor with the highest degree from non active actors) as our second seed actor (Table \ref{tab:toyNetMeasures}). Since actors 4, 5, and 2 are already active, we select actor 9. Thanks to that we are able to activate additional part of network resulting in 11 activated actors (100\% of network) after four iterations.}
 \label{fig:sq_example}
\end{figure*}

Most of the earlier influence maximisation research was based on the selection of all seeds at the beginning of the process, without additional actions taken after the process is launched. It was a different assumption than in real information spreading processes where various actions are taken during the process to improve its performance. For instance, an additional marketing budget is used, or new content is revealed. 

An example of such a process would be a \textit{The Grand Tour} show, where Amazon, instead of publishing all episodes at once (a traditional approach for VOD platforms like Netflix or Amazon Prime Video), decided to release episodes weekly to keep the audience for longer on their platform. Clearly, this led to an extension of Amazon Prime subscription period, but also led to a situation where people had more time to discover other content and, possibly, also started watching it. As a result, more of them decided to keep their their Prime memberships.

Recently, several attempts were taken to model additional seeding actions during spreading processes in the form of sequential seeding \cite{jankowski2017balancing}, adaptive seeding \cite{seeman2013adaptive}, seeding scheduling \cite{sela2015improving, goldenberg2018timing}, and active seeding \cite{sela2018active}. Sequential seeding research used the highest decomposition of the problem, starting from a single seed per stage (i.e. on seed per spreading model iteration) to evaluate the effect of spreading the seeding process over time. 

Several strategies were analysed. The first one was unconditional seeding which used a small number of seeds at the beginning of the process and then in each stage of simulation, additional seed or seeds were used \cite{jankowski2017balancing}. Another approach within the same study was based on revival mode. In this extension, instead of adding a new seed during each iteration, we add a new seed only when the activation cascade stops, thus we revive the spreading process. Experiments showed that using all seeds at the beginning is not the best strategy. A large fraction of nodes activated during the seeding stage can be activated in the natural spreading process by their neighbours. Saved seeding resources can be used to activate nodes difficult to reach with the natural spreading process, for example, in isolated network segments like in our toy examples (Fig.~\ref{fig:ss_example} and Fig.~\ref{fig:sq_example}). 




The performance of sequential seeding was analysed for the most often used heuristics like degree-based selection or greedy approach with better results for all seed selection strategies. Following study \cite{jankowski2017balancing} also verified the performance of seed selection with the use of Vote-Rank method \cite{zhang2016identifying} and top-k strategy for influence maximisation for networks with community structure \cite{he2015novel}. Another studies analysed the potential of seed selection methods based on entropy centralities~\cite{ni2019sequential} and the role of network typologies \cite{liu2018sequential}.

Unfortunately, all previous research was limited to simple one layer networks and recently temporal networks (tab. \ref{tab:landscape}). Thus, we have decided to extend previous work on sequential seeding to a multilayer scenario. 

\begin{table}
\centering
\begin{tabular}{l|r}
 One layer network & \cite{jankowski2017balancing, jankowski2017seeds, jankowski2018probing, jankowski2018strategic, ni2019sequential, liu2018sequential, seeman2013adaptive, sela2015improving, goldenberg2018timing, sela2018active} \\
 Temporal network & \cite{michalski2020effective} \\ 
 Multilayer multilayer & - \\ 
\end{tabular}
\caption{The summary of previous works on sequential seeding and similar approaches in relation to network on which they have been evaluated}
\label{tab:landscape}
\end{table}

Two approaches have been adapted. The first one is a classical sequential seeding (SQ). The toy example of this process is presented in Fig. \ref{fig:sq_example}. For SQ, first we select seeds the same way as for the single stage seeding (SS) process, but instead of activating all of them at once, we add one seed in each stage (one seed per iteration of ICM process) taking as a seed the node which is the highest ranked not activated node in our ranking list. We add seeds until we consume the whole seeding budget or there is no one else to activate (i.e., all nodes in the network are already active). As it can be seen in the presented toy examples for single stage (Fig. \ref{fig:ss_example}) and sequential (Fig. \ref{fig:sq_example}) seeding, using the second approach allowed us to activate an additional section of the network, increasing the final coverage of the process. 

The second approach was sequential seeding with revival (SQr) where instead of adding one seed in every stage, we wait until the spreading stops and only then add a new seed to revive the process.

\subsection{Coordinated execution}
\label{sec:coordinated}

Coordinated execution principle was initially designed for one layer networks and introduced in \cite{jankowski2018probing}. In our paper, we adjust it to the multilayer scenario. It allows us to evaluate and compare different seed activation strategies using Independent Cascade model despite the fact that ICM is not a deterministic model.

In the coordinated execution approach, instead of running ICM and drawing for each active actor, if it can activate its neighbour or not, we preselect the edges which can transmit the influence. To be more specific, for each network we create a number of instances of this network where for each edge, based on ICM propagation probability, we assign a binary choice independently for \textit{A} to \textit{B} and \textit{B} to \textit{A} telling us if \textit{A} can activate \textit{B} and vice versa.

Using this approach, for each network instance, we can easily compare the results for single stage seeding and sequential seeding since they are not influenced by the drawing results during ICM spreading. In other words, the spreading path will always be the same, e.g., if an actor \textit{A} is activated, it will always activate \textit{B} and never \textit{C} regardless of the seed activation strategy.

\section{Experiment setup}
\label{sec:experiment}
\begin{table}
\begin{tabular}{l r r r r p{4cm}}
 Name & Layers & Actors & Nodes & Edges & Description \\\hline
 N1 & 5 & 61 & 224 & 620 & AUCS CS-AARHUS \cite{rossi2015towards}\\
 N2 & 3 & 241 & 674 & 1370 & Ckm Physicians Innovation \cite{coleman1957diffusion} \\
 N3 & 37 & 417 & 2034 & 3588 & EU Air Transportation \cite{cardillo2013emergence} \\
 N4 & 3 & 71 & 212 & 1659 & Lazega Law Firm \cite{snijders2006new}\\\hline
 N5 & 2 & 1000 & 2000 & 5459 & \multirow{3}{4cm}{Each layer is an Erdős–Rényi network generated according to \cite{magnani2013formation} using multinet\cite{multinet} library}\\
 N6 & 3 & 1000 & 3000 & 7136 & \\
 N7 & 5 & 1000 & 5000 & 15109 & \\\hline
 N8 & 2 & 1000 & 2000 & 4223 & \multirow{3}{4cm}{Each layer is a scale-free network generated according to \cite{magnani2013formation} using multinet\cite{multinet} library }\\
 N9 & 3 & 1000 & 3000 & 5010 & \\
 N10 & 5 & 1000 & 5000 & 10181 & \\
\hline
\end{tabular}
\caption{Ten networks used in experiments, their parameters and short description.}
\label{tab:networks}
\end{table}

\begin{table}
\centering
\begin{tabular}{p{2cm} p{2.2cm} p{4cm} }
 Parameter & Values & Description \\ \hline
 N - Network & N1 - N10 & Four real networks (N1-N4) and six artificial (N4-N10). For details please see table \ref{tab:networks} \\ \hline
 PP - Propagation probability & 0.01, 0.02, 0.03, 0.05, 0.10, 0.20, 0.30, 0.40, 0.50 & Nine different values of propagation probability for Independent Cascade model \\ \hline
 SC - Seed count & 0.02, 0.05, 0.10, 0.20 & The percentage of the actors in the network selected as seeds \\ \hline
 3S - Seed selection strategy & degree centrality, neighbourhood size, random & Three seed selection strategies \\ \hline
 SA - Seed activation strategy & SS, SQ, SQr & Three seed distribution strategies. SS - single stage seeding, SQ - sequential seeding, SQr - sequential seeding with revival\\

\end{tabular}
\caption{Values for each parameter evaluated during experiments.}
\label{tab:parameteres}
\end{table}

The experiments have been performed using the \textit{multinet library} \cite{multinet} and ten multilayer networks (table \ref{tab:networks}). Following the coordinated execution principle \cite{jankowski2018probing} one hundred instances of each network for each propagation probability (PP) were generated by assigning binary choices of propagation or not for each edge, independently for \textit{A} to \textit{B} and \textit{B} to \textit{A} activation. This resulted in 9,000 ($10\times9\times100$) network instances. For each network instance, we have run 36 ($4\times3\times3$) simulations for all possible parameters combinations with four different seed counts, three seed selection strategies, and finally, three seed activation strategies (all parameters are described in table \ref{tab:parameteres}). In total, there were 324,000 different simulation cases.

\section{Results}
\label{sec:results}

\begin{figure*}
 \centering
 \includegraphics[width=\textwidth]{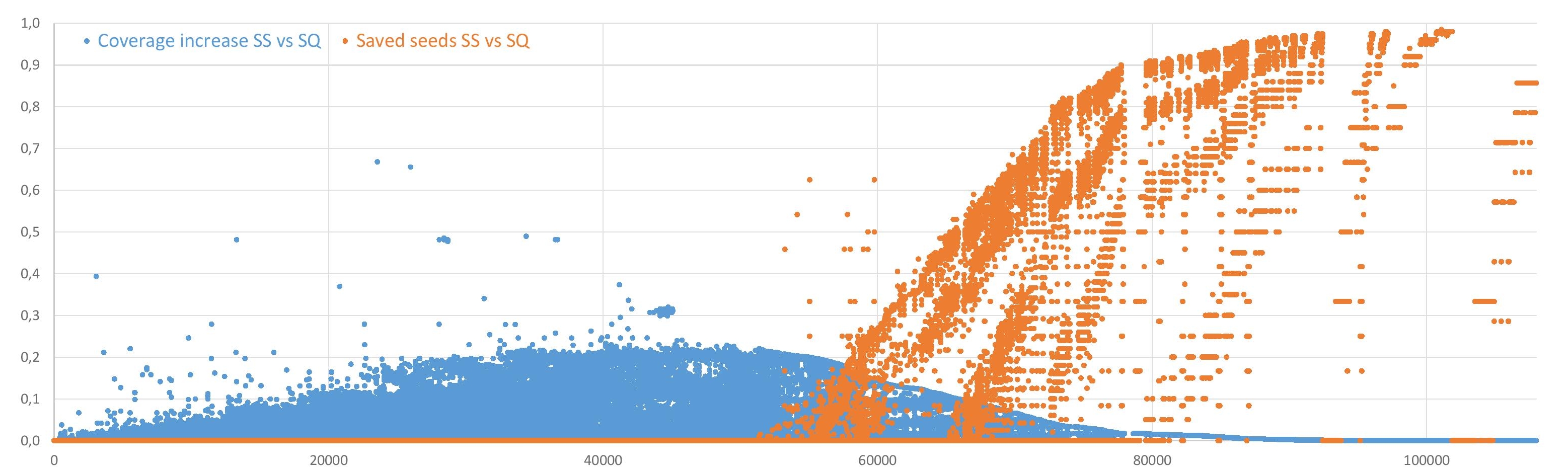}
 
 \caption{The comparison between SS and SQ strategy for all 108,000 cases. The results have been sorted by the SS percentage of activated actors. Blue colour indicates the gain i.e. how much more actors have been activated by SQ in compassion to SS. Orange colour indicates how big percentage of seeds have not been used (i.e. the budget saved by SQ strategy) after activating all actors in the network.}
 \label{fig:ss_vs_sq_gain}
\end{figure*}

\begin{figure*}
 \centering
 \includegraphics[width=\textwidth]{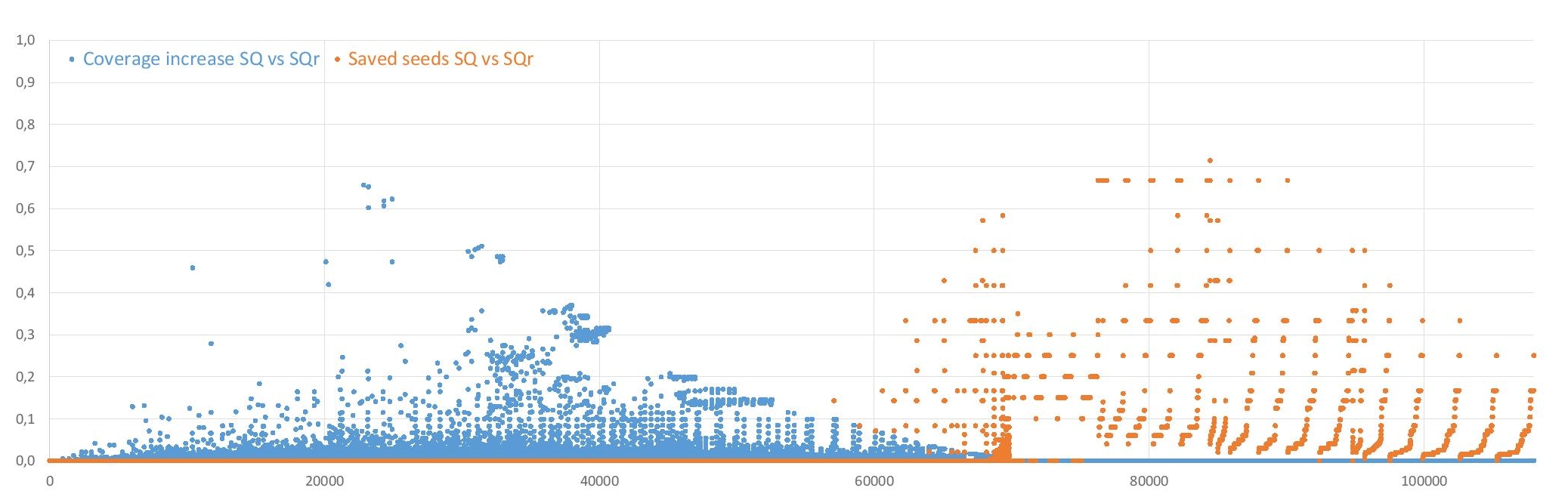}
 \caption{The comparison between SQ and SQr strategy for all 108,000 cases. The results have been sorted by the SS percentage of activated actors. Blue colour indicates the increase in the coverage in the percentage of all actors in the network i.e. how much more actors have been activated by SQ in compassion to SS. Orange colour indicates how big percentage of seeds have not been used (i.e. the budget saved by SQ strategy) after activating all actors in the network.}
 \label{fig:sq_vs_SQr_gain}
\end{figure*}

\begin{figure*}
 \centering
 \includegraphics[width=\textwidth]{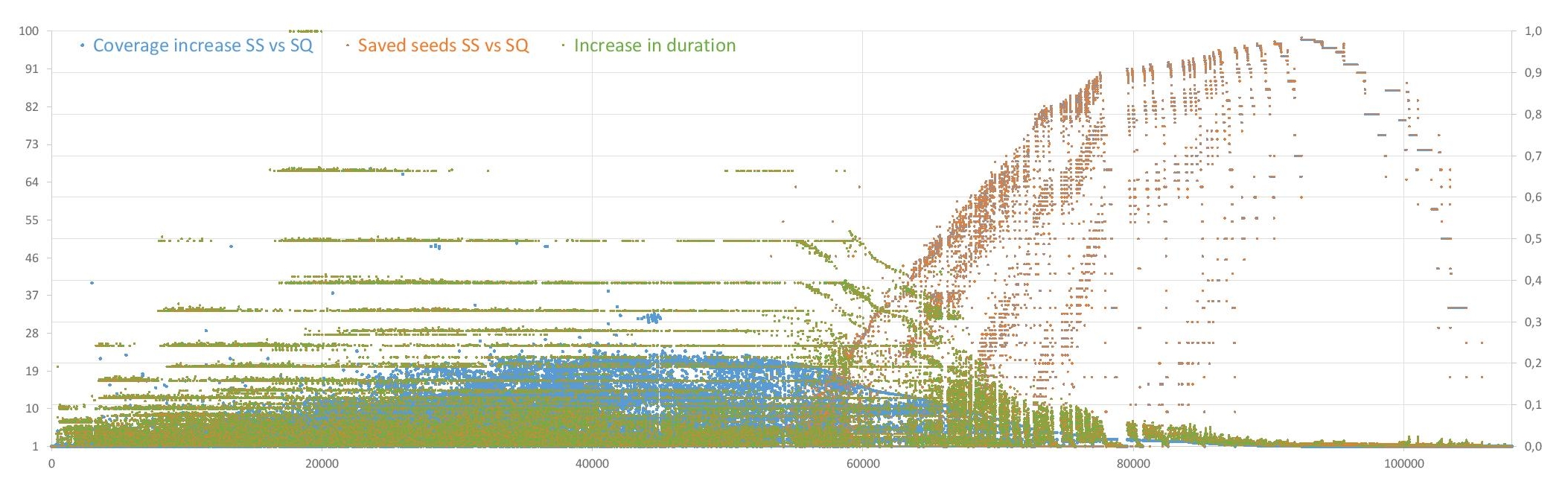}
 \caption{The comparison between 108,000 cases of SS and SQ strategy. In the background and on the right Y-ax we have the same information as on figure \ref{fig:ss_vs_sq_gain} while on the front and left Y-ax we have the information on how many times more iterations SQ needed to finish in comparison to SS.}
 \label{fig:ss_vs_sq_time}
\end{figure*}

In the figure \ref{fig:ss_vs_sq_gain} the comparison between 108,000 cases of single stage (SS) and sequential (SQ) seeding strategies is presented. The results have been sorted by the SS percentage of activated actors. Blue colour indicates the increase in the coverage in the percentage of all actors in the network, i.e., how much more actors have been activated by SQ in compassion to SS. Orange colour indicates how big percentage of seeds have not been used after activating all actors in the network (i.e., the budget saved by SQ strategy). 

Sequential seeding always achieves at least the same results as SS, and in 74\% of the cases, SQ performs better than SS. What is more, in 31.75\% of the cases (including 43.63\% of cases where SQ yields the same results as SS) SQ does not have to use all seeds to activate all actors in the network. This shows the new benefit of the sequential seeding approach, i.e., by observing the process as it progresses, by wiser decisions we can save some portion of the budget needed to activate seeds. This feature might be crucial if one is doing an advertisement campaign, and each seed is an actual cost for the company. What is more, previous papers investigating sequential seeding in simple one layer networks looked only at the increase in coverage or the interplay between coverage in time, but none of them evaluated seeds savings.

On average, SQ was able to activate 1.074 times more nodes than SS, with the minimum one (i.e., the same number) and the maximum nine times more. Regarding the saved seeds, on average SQ saved 21\% of seeds with 0\% as minimum and 95.5\% as maximum.

Similar but not so overwhelming results can be noticed when we compare SQ with sequential seed with revival (SQr - see section \ref{sec:sq}). It is always at least as good as SQ, in 44.21\% of cases it is better than SQ, and in 1.32\% of cases SQr was able to activate all nodes in the network with lower number of seeds than SQ (Fig. \ref{fig:sq_vs_SQr_gain}).

However, the higher coverage and savings of SQ and SQr comes with the price. Similarly to the observations presented in \cite{jankowski2017balancing}, also in the case of multilayer networks using SQ and SQr seed activation strategy results in a much longer spreading process. On average, for SQ it takes nine times and for SQr 12.9 times more iterations of the ICM to finish the spreading process (Fig. \ref{fig:ss_vs_sq_time}).

In the following sections, we will take a more in-depth look at each of those three main elements, i.e., spread coverage, spread duration, and saved seeds, in the context of the parameters we have used in the experiments.

\begin{table}[!ht]
\begin{tabular}{l | r r r | r r | r r}
Parameter &\multicolumn{3}{c|}{Spread coverage} & \multicolumn{2}{c|}{Saved seeds} & \multicolumn{2}{c}{Spread duration}\\
&SS&SQ&SQr&SQ&SQr&SQ&SQr\\\hline

\multicolumn{8}{l}{Network}\\\hline
N1&71\%&73\%&75\%&13\%&25\%&151\%&216\%\\
N2&58\%&62\%&65\%&15\%&20\%&400\%&742\%\\
N3&88\%&89\%&90\%&27\%&40\%&131\%&172\%\\
N4&61\%&67\%&68\%&6\%&6\%&751\%&973\%\\\hline
N5&63\%&65\%&65\%&32\%&36\%&1048\%&1673\%\\
N6&67\%&70\%&70\%&37\%&41\%&882\%&1446\%\\
N7&81\%&84\%&84\%&51\%&56\%&570\%&953\%\\
N8&51\%&57\%&57\%&0\%&0\%&2108\%&2719\%\\
N9&56\%&62\%&63\%&3\%&3\%&1903\%&2503\%\\
N10&73\%&77\%&77\%&27\%&29\%&1070\%&1526\%\\\hline

\multicolumn{8}{l}{Seed Selection Strategy}\\\hline
Degree&67\%&71\%&72\%&21\%&26\%&956\%&1403\%\\
Neigbourhood&67\%&71\%&72\%&21\%&26\%&952\%&1396\%\\
Random&67\%&70\%&70\%&21\%&26\%&796\%&1077\%\\\hline

\multicolumn{8}{l}{Propagation Probability}\\\hline
0.01&16\%&18\%&18\%&0\%&0\%&1666\%&2402\%\\
0.02&29\%&33\%&34\%&0\%&0\%&1205\%&2075\%\\
0.03&41\%&46\%&47\%&0\%&1\%&1027\%&1830\%\\
0.05&59\%&65\%&67\%&4\%&7\%&873\%&1454\%\\
0.1&81\%&87\%&88\%&14\%&19\%&794\%&1023\%\\
0.2&91\%&95\%&96\%&32\%&40\%&670\%&770\%\\
0.3&94\%&97\%&97\%&42\%&50\%&620\%&693\%\\
0.4&95\%&97\%&98\%&47\%&55\%&621\%&686\%\\
0.5&95\%&98\%&98\%&49\%&58\%&635\%&696\%\\\hline

\multicolumn{8}{l}{Seed Count}\\\hline
0.02&61\%&63\%&64\%&8\%&12\%&234\%&391\%\\
0.05&65\%&68\%&68\%&16\%&23\%&489\%&773\%\\
0.1&68\%&73\%&73\%&25\%&30\%&952\%&1394\%\\
0.2&72\%&80\%&80\%&35\%&38\%&1930\%&2611\%\\\hline

\end{tabular}
\caption{Influence of experiments parameters on coverage, saved seeds and time lost. The results for each parameter have been averaged for all other parameters}
\label{tab:results_all_param}
\end{table}

\subsection{Network type}
When we look at the results for different networks (Table \ref{tab:results_all_param}) we can see the number of layers is affecting the final coverage, what is in line with previous research on that topic. However, there is no evidence that SQ or SQr improvement in terms of coverage depends on the network source (real vs synthetic) or type (random vs scale-free). 

However, the number of layers seems to affect saved seeds. Networks with a higher number of layers (N3, N7 and N10) have on average the highest number of saved seeds in their groups (real networks, synthetic-random, synthetic-scale-free respectively). 

At the same time, the increase in the number of iterations is smaller for networks with higher number of layers. What is also interesting, the increased spread duration does not affect real networks as much as synthetic networks, which is very good news for real applications. 

\subsection{Seed selection strategy}
There is no evidence that the seed selection strategy affects the influence of SQ or SQr over the spreading process (Table \ref{tab:results_all_param}). All three approaches yield similar results. 

It needs to be emphasised that this is an expected outcome, as sequential seeding was intended to work on top of, and be independent on, any seed selection strategy, and our results confirm that.

\subsection{Propagation probability}
SQ and SQr improve the total coverage regardless of the propagation probability (PP) used in ICM (Table \ref{tab:results_all_param}). They are of course affected by it in a similar way as single stage seeding, i.e., smaller the PP is, the smaller coverage we have. 

For higher PP (PP $\geqslant$ 0.1) where SQ and SQr cannot improve the coverage too much, since usually the entire network is activated, they allow to save the seeding budget. As mentioned before, this is a powerful advantage of sequential seeding where by observing the process we can decide to stop acquiring additional seeds and save our budget.

In terms of process duration, SQ and SQr affect more spreads where we have smaller propagation probability. However, this could also be caused by the simple fact that SQ and SQr ended "earlier" because there was no one else to activate.

\subsection{Seed count}

The last parameter we have investigated is seed count (SC). In this case, the results are very intuitive (Table \ref{tab:results_all_param}). 
The higher number of seeds results in higher coverage for SS, SQ and SQr. The bigger seeding budget we have, the more effective SQ and SQr are, and allow for a higher increase in the final coverage.

For all values of SC, we can observe saved seeds, with a higher number of saved seeds for higher values of SC. However, this might be caused by the fact that the bigger budget we have at the beginning, the more we can save at the end.

Finally, the more seeds we have the budget for, the longer lasts the spreading process. Because we acquire only one seed per stage, our seeding is usually much longer than the entire spreading in the case of SS. For example, for network N9, SS ended on average after 5-6 iterations regardless of SC, while just seeding process for SQ took on average 20, 50, 100 and 178 iterations for SC 0.02, 0.05, 0.1, 0.2 respectively. Please note that the lower number of iterations than the number of seeds in the case of SC $=$ 0.2 is a result of the process ending before seeding ended due to activation of all nodes in the network. 

\subsection{Statistical analysis}

The last part of our analysis was the statistical significance evaluation of the results. To do so, Wilcoxon signed-rank test was used as a measure of the difference between sequential and single stage seeding. Wilcoxson signed rank is a nonparametric counterpart of the paired t-test and often used for algorithms comparison \cite{hassan2020operational} since it does not require normal distribution. The results presented in Table~\ref{tab:statistics} show higher coverage in terms of Hodges–Lehmann estimator $\Delta$ for sequential seeding when compared to single stage seeding with different values of used parameters. Overall results from all simulations showed $\Delta$ = 3.45 with p-value < 2.2e-16. Values of $\Delta$ > 0 confirm significantly higher values for coverage of sequential seeding when compared to the single stage approach. For more detailed results, Hodges-Lehmann estimator was computed for each propagation probability PP used in simulations. It confirms the higher performance of the proposed approach (p-value < 0.05) for all used propagation probabilities. The highest performance ($\Delta$ = 6.39) was observed for the propagation probability at the medium level PP = 0.05 while the lowest performance ($\Delta$ = 1.14) for low PP = 0.01. In general, lower performance if observed for low and high propagation probability values. For high propagation probability, any used strategy can bring good results while for low PP any strategy can fail. 

The analysis performed in terms of seed count shows the best performance of sequential seeding for the highest seed count (0.2); $\Delta$ = 7.80. The lowest performance ($\Delta$ = 1.16) was observed for the lowest seed count, i.e., 0.02. The main reason is the fact that together with the high number of seeds, there is a higher chance that we will select as seed nodes that will be activated by their neighbours anyway. Sequential seeding reduces this problem by the activation of additional seeds when the process stops. 

Another dimension of the analysis took into account seed selection strategy. Degree and Neighbourhood based seed selection delivered similar results in terms of $\Delta$ with values 4.20 and 4.07, respectively. The lowest performance ($\Delta$ = 2.34) was observed for random seed selection. 

The results depended on the used networks. For real networks, the highest difference with $\Delta$ = 5.40 was obtained for N3 (EUAir) network. The lowest performance with $\Delta$ = 1.93 sequential seeding achieved for N1 (AUCS) network. Even though the overall performance is at different levels for all networks, sequential seeding always outperforms single stage seeding with p-value < 0.05. Differences were also observed for synthetic networks. For random networks (N5-N7) a growing number of layers resulted in increased performance of sequential seeding with $\Delta$ up to 3.28 for the network with five layers, while for scale-free networks (N8-N10) a lower number of layers delivered better results at the level of 5.35 and 5.65 for two and three layers respectively. 

\begin{table}
\begin{tabular}{l r r r }
 Parameter & Value & Hodges – Lehmann $\Delta$ & p-value \\\hline
 All & All & 3.45 & < 2.2e-16 \\\hline
 
PP & 0.01 & 1.14 & < 2.2e-16 \\
& 0.02 & 3.11 & < 2.2e-16 \\
& 0.03 & 4.80 & < 2.2e-16 \\
& 0.05 & 6.39 & < 2.2e-16 \\
& 0.1 & 5.63 & < 2.2e-16 \\
& 0.2 & 3.05& < 2.2e-16 \\
& 0.3 & 2.47 & < 2.2e-16 \\
& 0.4 & 2.37 & < 2.2e-16 \\
& 0.5 & 2.29 & < 2.2e-16 \\\hline

 SC & 0.02 & 1.16 & < 2.2e-16 \\ 
 & 0.05 & 2.47 & < 2.2e-16 \\
 & 0.1 & 4.85 & < 2.2e-16 \\
 & 0.2 & 7.80 & < 2.2e-16 \\\hline

3S & degree & 4.20 & < 2.2e-16 \\ 
 & neigbourhood & 4.07 & < 2.2e-16 \\
 & random & 2.34 & < 2.2e-16 \\\hline
 
 Real nets & N1 & 1.93 & < 3.66e-14 \\ 
 & N2 & 3.17 & < 2.2e-16 \\
 & N3 & 5.40 & < 2.2e-16 \\
 & N4 & 2.28 & 1.702e-08 \\\hline

 Synthetic nets & N5 & 1.58 & < 2.2e-16 \\
 & N6 & 1.35 & < 2.2e-16 \\
 & N7 & 3.28 & 1.654e-13 \\
 & N8 & 5.35 & < 2.2e-16 \\
 & N9 & 5.65 & < 2.2e-16 \\
 & N10 & 3.45 & < 2.2e-16 \\\hline

\end{tabular}
\caption{The differences between results for single stage and sequential seeding represented by Hodges–Lehmann estimator $\Delta$ with positive values for better results for sequential seeding.}
\label{tab:statistics}
\end{table}

\subsection{arXiv network science network}

In this section, we aim at evaluating if the results for networks N1-N10 will scale up for the larger network presented in table~\ref{tab:networksL}. Due to time and equipment limitation, the experiments were limited to three propagation probability values $PP\in\{0.05, 0.1, 0.2\}$, the rest of parameters is the same as described in table~\ref{tab:parameteres}.

\begin{table}
\begin{tabular}{l r r r r p{3cm}}
 Name & Layers & Actors & Nodes & Edges & Description \\\hline
 N11 & 13 & 14,065 & 26,796 & 59,026 & arXiv netscience\cite{de2015identifying}\\
 \hline
\end{tabular}
\caption{Large network used in additional experiments}
\label{tab:networksL}
\end{table}
 
The results confirm the advantages and disadvantages of sequential seeding over single stage seeding for networks N1-N10.

The SQ was better than SS in 100\% cases, with SQ being able to activate on average 25\% more actors than SS. 
The SQr was better than SQ in 96\% of cases, but the average increase in the number of activated actors was just 1\%.

With this network and selected propagation probabilities, we have not observed any saved seeds since the maximum number of activated actors were 71.2\% for SS, 89.6\% for SQ and 89.7\% for SQr. Thus, if no simulation was able to activate all actors, no seed savings were possible.

Finally, it took SQ on average almost 17 times and for SQr 28 times longer than SS to finish.

\section{Conclusions}
Multilayer networks are usually considered as a better approximation of real interactions between nodes in the network (especially in the case of social networks) since they allow to model multiple relations between nodes. Unfortunately, most of the existing research is focused on simple one layer networks, and the same problem was with the sequential seeding approach. In this paper, we have addressed this issue by extending sequential seeding to a multilayer scenario and evaluating it on four real and six synthetic networks. The main results are following: \textit{(i)} sequential seeding is always at least as good as single stage and in many cases (74\%) is better; \textit{(ii)} sequential seeding very often (31.75\%) allows to save seeding budget since it does not need so many seeds to activate all nodes in the network (this is especially important in cases where sequential and single stage seeding produce the same results); however \textit{(iii)} better coverage and saved seeding budget comes with the price of extended duration of the spreading campaign (on average 9 times longer process). 

When looking at the results in detail, we can observe that the number of layers is bound to the savings of seeds: the more layers the network has, the more budget is saved. Another phenomenon observed is that sequential seeding performs best for rather small propagation probabilities, but for higher it can also contribute to saving budget, so multiple real-life scenarios can demonstrate the superiority of this method, but the outcome can be of a different kind. Lastly, the sequential seeding method also proved that it is independent from the underlying seed selection strategy.

Due to longer spreading duration, sequential seeding should be used with caution in campaigns where we have a fixed deadline for influencing people, e.g., during a presidential campaign. On the other hand, this approach is useful in applications where we do not care so much about time, have limited resources, but want to influence as many people as possible.

In terms of practical applications, if there exists a group of highly connected users, it would be enough to seed a small fraction of them instead of spending a larger budget. Information will flow in a natural way and the saved budget can be used in other network areas. Similar solution was described in Facebook study\cite{iyer2018costs} where they investigate the costs of overambitious seeding of social products. Authors showed the advantages of spreading seeds over time and proposed gradual seeding based partially on our work\cite{jankowski2017balancing} on sequential seeding in one layer networks.

The final issue we would like to discuss is if the results of this study are specific solely to the ICM or can be generalised to other spreading models. Based on our previous works (see section~\ref{sec:sq}) and the formal proof of the non-decreasing coverage we have included in\cite{jankowski2018probing}, regardless of the influence model, sequential seeding will yield at least the same results as single stage seeding if the seeds are selected based on some ranking.

We can consider the Linear Threshold Model\cite{kempe2003maximizing} (LTM) as an example. Assuming we have a budget for ten seeds and select the top ten nodes according to degree centrality, we will get at least the same coverage regardless if we activate all of them in a single stage, at the beginning, or activate them in a sequence over the first ten iterations. However, by activating seeds in a sequence, we have a chance that some of these top ten seeds will be activated by the previously activated ones. If this happens, we save a part of seeding budget and we are able to activate the next inactive node in a ranking which will become a new seed. This will lead to a longer spreading process, similarly to the ICM case. The same argument can be made for any threshold-based model and as we mentioned above, any seed set based model with two states. The issue is more complicated with equation-based models, like the Bass model\cite{bass1969new}, where sequential seeding cannot be used directly. However, if we use agent-based Bass models\cite{rand2011agent,chica2017building} then the advantages and disadvantages of using SQ should be the same as for ICM.

Of course, to assess the impact of sequential sequential seeding in a real network scenario, additional studies are required. Some of them, for LTM in one layer networks and sequential seeding like methods, have been already done\cite{sela2018active, goldenberg2018timing}. Summing up, sequential seeding will yield the same results as single stage seeding, but the scale of possible performance gain depends on the network configuration and social influence model, when we consider a family of models in which the total spread is a monotonically increasing function.

The models where sequential seeding might not produce the same results are more complex models of contagion. For instance, these can be three-state models like SIR, UAF\cite{scata2016impact}, models where the node can go back to the previous state like SIS, UAU\cite{zang2018effects} or Voter model\cite{holley1975ergodic}. For example, if we use sequential seeding to increase peoples' awareness (transition from \textit{unaware} to \textit{aware}), then the increased duration of that process allows for more chances for the node to switch to another state (\textit{aware} to \textit{forgot}) or switch back (\textit{aware} to \textit{unaware}). Thus, the effectiveness of sequential seeding might be limited to some cases where the delay is not big enough to nullify the gain from SQ. However, further research is needed to fully understand the interplay between those processes in complex models of contagion.

When considering future work directions, apart from the evaluation of different spreading models mentioned above, one of the most interesting for us is to observe how sequential seeding performs when the types of layers are of different kinds. One can also think of varying the propagation probability for each layer reflecting different intensities of interactions, but this requires additional experiments as well as interpretations coming from the social science field.


%
%

%

\begin{acknowledgments}
This work has been supported by National Science Center, Poland, grant number 2016/21/B/HS4/01562.
\end{acknowledgments}

\section*{Data and code availability statement}
All real networks are available at CoMuNe lab repository (\url{https://manliodedomenico.com/data.php}). Apart from that, both real and synthetic networks are published at GitHub repository\cite{brodka2021github}, FullNet folder).

To ease the process of reproducing the experiment, we also created a CodeOcean capsule~\cite{brodka2021sequentialCode} with the code used for experiments and the AUCS network as an exemplary dataset, but other networks can be also evaluated with this code when taken from CoMuNe lab repository or GitHub (see above).

Additionally, for the sake of reproducibility, all networks generated by the coordinated execution procedure (900 networks for each evaluated network), the results of all experiments as well as the R code, were published at GitHub repository\cite{brodka2021github} as well.

\section*{REFERENCES}
\bibliography{main}

\providecommand{\noopsort}[1]{}\providecommand{\singleletter}[1]{#1}%
\begin{thebibliography}{47}%
\makeatletter
\providecommand \@ifxundefined [1]{%
 \@ifx{#1\undefined}
}%
\providecommand \@ifnum [1]{%
 \ifnum #1\expandafter \@firstoftwo
 \else \expandafter \@secondoftwo
 \fi
}%
\providecommand \@ifx [1]{%
 \ifx #1\expandafter \@firstoftwo
 \else \expandafter \@secondoftwo
 \fi
}%
\providecommand \natexlab [1]{#1}%
\providecommand \enquote  [1]{``#1''}%
\providecommand \bibnamefont  [1]{#1}%
\providecommand \bibfnamefont [1]{#1}%
\providecommand \citenamefont [1]{#1}%
\providecommand \href@noop [0]{\@secondoftwo}%
\providecommand \href [0]{\begingroup \@sanitize@url \@href}%
\providecommand \@href[1]{\@@startlink{#1}\@@href}%
\providecommand \@@href[1]{\endgroup#1\@@endlink}%
\providecommand \@sanitize@url [0]{\catcode `\\12\catcode `\$12\catcode
  `\&12\catcode `\#12\catcode `\^12\catcode `\_12\catcode `\%12\relax}%
\providecommand \@@startlink[1]{}%
\providecommand \@@endlink[0]{}%
\providecommand \url  [0]{\begingroup\@sanitize@url \@url }%
\providecommand \@url [1]{\endgroup\@href {#1}{\urlprefix }}%
\providecommand \urlprefix  [0]{URL }%
\providecommand \Eprint [0]{\href }%
\providecommand \doibase [0]{http://dx.doi.org/}%
\providecommand \selectlanguage [0]{\@gobble}%
\providecommand \bibinfo  [0]{\@secondoftwo}%
\providecommand \bibfield  [0]{\@secondoftwo}%
\providecommand \translation [1]{[#1]}%
\providecommand \BibitemOpen [0]{}%
\providecommand \bibitemStop [0]{}%
\providecommand \bibitemNoStop [0]{.\EOS\space}%
\providecommand \EOS [0]{\spacefactor3000\relax}%
\providecommand \BibitemShut  [1]{\csname bibitem#1\endcsname}%
\let\auto@bib@innerbib\@empty
\bibitem [{\citenamefont {Kempe}, \citenamefont {Kleinberg},\ and\
  \citenamefont {Tardos}(2003)}]{kempe2003maximizing}%
  \BibitemOpen
  \bibfield  {author} {\bibinfo {author} {\bibfnamefont {D.}~\bibnamefont
  {Kempe}}, \bibinfo {author} {\bibfnamefont {J.}~\bibnamefont {Kleinberg}}, \
  and\ \bibinfo {author} {\bibfnamefont {{\'E}.}~\bibnamefont {Tardos}},\
  }\bibfield  {title} {\enquote {\bibinfo {title} {Maximizing the spread of
  influence through a social network},}\ }in\ \href@noop {} {\emph {\bibinfo
  {booktitle} {Proceedings of the ninth ACM SIGKDD international conference on
  Knowledge discovery and data mining}}}\ (\bibinfo {year} {2003})\ pp.\
  \bibinfo {pages} {137--146}\BibitemShut {NoStop}%
\bibitem [{\citenamefont {Jankowski}\ \emph
  {et~al.}(2017{\natexlab{a}})\citenamefont {Jankowski}, \citenamefont
  {Br{\'o}dka}, \citenamefont {Kazienko}, \citenamefont {Szymanski},
  \citenamefont {Michalski},\ and\ \citenamefont
  {Kajdanowicz}}]{jankowski2017balancing}%
  \BibitemOpen
  \bibfield  {author} {\bibinfo {author} {\bibfnamefont {J.}~\bibnamefont
  {Jankowski}}, \bibinfo {author} {\bibfnamefont {P.}~\bibnamefont
  {Br{\'o}dka}}, \bibinfo {author} {\bibfnamefont {P.}~\bibnamefont
  {Kazienko}}, \bibinfo {author} {\bibfnamefont {B.~K.}\ \bibnamefont
  {Szymanski}}, \bibinfo {author} {\bibfnamefont {R.}~\bibnamefont
  {Michalski}}, \ and\ \bibinfo {author} {\bibfnamefont {T.}~\bibnamefont
  {Kajdanowicz}},\ }\bibfield  {title} {\enquote {\bibinfo {title} {Balancing
  speed and coverage by sequential seeding in complex networks},}\ }\href@noop
  {} {\bibfield  {journal} {\bibinfo  {journal} {Scientific reports}\ }\textbf
  {\bibinfo {volume} {7}},\ \bibinfo {pages} {1--11} (\bibinfo {year}
  {2017}{\natexlab{a}})}\BibitemShut {NoStop}%
\bibitem [{\citenamefont {Jankowski}\ \emph
  {et~al.}(2018{\natexlab{a}})\citenamefont {Jankowski}, \citenamefont
  {Szymanski}, \citenamefont {Kazienko}, \citenamefont {Michalski},\ and\
  \citenamefont {Br{\'o}dka}}]{jankowski2018probing}%
  \BibitemOpen
  \bibfield  {author} {\bibinfo {author} {\bibfnamefont {J.}~\bibnamefont
  {Jankowski}}, \bibinfo {author} {\bibfnamefont {B.~K.}\ \bibnamefont
  {Szymanski}}, \bibinfo {author} {\bibfnamefont {P.}~\bibnamefont {Kazienko}},
  \bibinfo {author} {\bibfnamefont {R.}~\bibnamefont {Michalski}}, \ and\
  \bibinfo {author} {\bibfnamefont {P.}~\bibnamefont {Br{\'o}dka}},\ }\bibfield
   {title} {\enquote {\bibinfo {title} {Probing limits of information spread
  with sequential seeding},}\ }\href@noop {} {\bibfield  {journal} {\bibinfo
  {journal} {Scientific reports}\ }\textbf {\bibinfo {volume} {8}},\ \bibinfo
  {pages} {1--9} (\bibinfo {year} {2018}{\natexlab{a}})}\BibitemShut {NoStop}%
\bibitem [{\citenamefont {Kivel{\"a}}\ \emph {et~al.}(2014)\citenamefont
  {Kivel{\"a}}, \citenamefont {Arenas}, \citenamefont {Barthelemy},
  \citenamefont {Gleeson}, \citenamefont {Moreno},\ and\ \citenamefont
  {Porter}}]{kivela2014multilayer}%
  \BibitemOpen
  \bibfield  {author} {\bibinfo {author} {\bibfnamefont {M.}~\bibnamefont
  {Kivel{\"a}}}, \bibinfo {author} {\bibfnamefont {A.}~\bibnamefont {Arenas}},
  \bibinfo {author} {\bibfnamefont {M.}~\bibnamefont {Barthelemy}}, \bibinfo
  {author} {\bibfnamefont {J.~P.}\ \bibnamefont {Gleeson}}, \bibinfo {author}
  {\bibfnamefont {Y.}~\bibnamefont {Moreno}}, \ and\ \bibinfo {author}
  {\bibfnamefont {M.~A.}\ \bibnamefont {Porter}},\ }\bibfield  {title}
  {\enquote {\bibinfo {title} {Multilayer networks},}\ }\href@noop {}
  {\bibfield  {journal} {\bibinfo  {journal} {Journal of complex networks}\
  }\textbf {\bibinfo {volume} {2}},\ \bibinfo {pages} {203--271} (\bibinfo
  {year} {2014})}\BibitemShut {NoStop}%
\bibitem [{\citenamefont {Michalski}, \citenamefont {Kazienko},\ and\
  \citenamefont {Jankowski}(2013)}]{michalski2013convince}%
  \BibitemOpen
  \bibfield  {author} {\bibinfo {author} {\bibfnamefont {R.}~\bibnamefont
  {Michalski}}, \bibinfo {author} {\bibfnamefont {P.}~\bibnamefont {Kazienko}},
  \ and\ \bibinfo {author} {\bibfnamefont {J.}~\bibnamefont {Jankowski}},\
  }\bibfield  {title} {\enquote {\bibinfo {title} {Convince a dozen more and
  succeed--the influence in multi-layered social networks},}\ }in\ \href@noop
  {} {\emph {\bibinfo {booktitle} {2013 International Conference on
  Signal-Image Technology \& Internet-Based Systems}}}\ (\bibinfo
  {organization} {IEEE},\ \bibinfo {year} {2013})\ pp.\ \bibinfo {pages}
  {499--505}\BibitemShut {NoStop}%
\bibitem [{\citenamefont {Boccaletti}\ \emph {et~al.}(2014)\citenamefont
  {Boccaletti}, \citenamefont {Bianconi}, \citenamefont {Criado}, \citenamefont
  {Del~Genio}, \citenamefont {G{\'o}mez-Gardenes}, \citenamefont {Romance},
  \citenamefont {Sendina-Nadal}, \citenamefont {Wang},\ and\ \citenamefont
  {Zanin}}]{boccaletti2014structure}%
  \BibitemOpen
  \bibfield  {author} {\bibinfo {author} {\bibfnamefont {S.}~\bibnamefont
  {Boccaletti}}, \bibinfo {author} {\bibfnamefont {G.}~\bibnamefont
  {Bianconi}}, \bibinfo {author} {\bibfnamefont {R.}~\bibnamefont {Criado}},
  \bibinfo {author} {\bibfnamefont {C.~I.}\ \bibnamefont {Del~Genio}}, \bibinfo
  {author} {\bibfnamefont {J.}~\bibnamefont {G{\'o}mez-Gardenes}}, \bibinfo
  {author} {\bibfnamefont {M.}~\bibnamefont {Romance}}, \bibinfo {author}
  {\bibfnamefont {I.}~\bibnamefont {Sendina-Nadal}}, \bibinfo {author}
  {\bibfnamefont {Z.}~\bibnamefont {Wang}}, \ and\ \bibinfo {author}
  {\bibfnamefont {M.}~\bibnamefont {Zanin}},\ }\bibfield  {title} {\enquote
  {\bibinfo {title} {The structure and dynamics of multilayer networks},}\
  }\href@noop {} {\bibfield  {journal} {\bibinfo  {journal} {Physics Reports}\
  }\textbf {\bibinfo {volume} {544}},\ \bibinfo {pages} {1--122} (\bibinfo
  {year} {2014})}\BibitemShut {NoStop}%
\bibitem [{\citenamefont {De~Domenico}\ \emph {et~al.}(2016)\citenamefont
  {De~Domenico}, \citenamefont {Granell}, \citenamefont {Porter},\ and\
  \citenamefont {Arenas}}]{de2016physics}%
  \BibitemOpen
  \bibfield  {author} {\bibinfo {author} {\bibfnamefont {M.}~\bibnamefont
  {De~Domenico}}, \bibinfo {author} {\bibfnamefont {C.}~\bibnamefont
  {Granell}}, \bibinfo {author} {\bibfnamefont {M.~A.}\ \bibnamefont {Porter}},
  \ and\ \bibinfo {author} {\bibfnamefont {A.}~\bibnamefont {Arenas}},\
  }\bibfield  {title} {\enquote {\bibinfo {title} {The physics of spreading
  processes in multilayer networks},}\ }\href@noop {} {\bibfield  {journal}
  {\bibinfo  {journal} {Nature Physics}\ }\textbf {\bibinfo {volume} {12}},\
  \bibinfo {pages} {901--906} (\bibinfo {year} {2016})}\BibitemShut {NoStop}%
\bibitem [{\citenamefont {Br{\'o}dka}, \citenamefont {Musial},\ and\
  \citenamefont {Jankowski}(2020)}]{brodka2020interacting}%
  \BibitemOpen
  \bibfield  {author} {\bibinfo {author} {\bibfnamefont {P.}~\bibnamefont
  {Br{\'o}dka}}, \bibinfo {author} {\bibfnamefont {K.}~\bibnamefont {Musial}},
  \ and\ \bibinfo {author} {\bibfnamefont {J.}~\bibnamefont {Jankowski}},\
  }\bibfield  {title} {\enquote {\bibinfo {title} {Interacting spreading
  processes in multilayer networks: a systematic review},}\ }\href@noop {}
  {\bibfield  {journal} {\bibinfo  {journal} {IEEE Access}\ }\textbf {\bibinfo
  {volume} {8}},\ \bibinfo {pages} {10316--10341} (\bibinfo {year}
  {2020})}\BibitemShut {NoStop}%
\bibitem [{\citenamefont {Basaras}\ \emph {et~al.}(2017)\citenamefont
  {Basaras}, \citenamefont {Iosifidis}, \citenamefont {Katsaros},\ and\
  \citenamefont {Tassiulas}}]{basaras2017identifying}%
  \BibitemOpen
  \bibfield  {author} {\bibinfo {author} {\bibfnamefont {P.}~\bibnamefont
  {Basaras}}, \bibinfo {author} {\bibfnamefont {G.}~\bibnamefont {Iosifidis}},
  \bibinfo {author} {\bibfnamefont {D.}~\bibnamefont {Katsaros}}, \ and\
  \bibinfo {author} {\bibfnamefont {L.}~\bibnamefont {Tassiulas}},\ }\bibfield
  {title} {\enquote {\bibinfo {title} {Identifying influential spreaders in
  complex multilayer networks: A centrality perspective},}\ }\href@noop {}
  {\bibfield  {journal} {\bibinfo  {journal} {IEEE Transactions on Network
  Science and Engineering}\ }\textbf {\bibinfo {volume} {6}},\ \bibinfo {pages}
  {31--45} (\bibinfo {year} {2017})}\BibitemShut {NoStop}%
\bibitem [{\citenamefont {Huang}\ \emph {et~al.}(2020)\citenamefont {Huang},
  \citenamefont {Chen}, \citenamefont {Wang},\ and\ \citenamefont
  {Ren}}]{huang2020identifying}%
  \BibitemOpen
  \bibfield  {author} {\bibinfo {author} {\bibfnamefont {X.}~\bibnamefont
  {Huang}}, \bibinfo {author} {\bibfnamefont {D.}~\bibnamefont {Chen}},
  \bibinfo {author} {\bibfnamefont {D.}~\bibnamefont {Wang}}, \ and\ \bibinfo
  {author} {\bibfnamefont {T.}~\bibnamefont {Ren}},\ }\bibfield  {title}
  {\enquote {\bibinfo {title} {Identifying influencers in social networks},}\
  }\href@noop {} {\bibfield  {journal} {\bibinfo  {journal} {Entropy}\ }\textbf
  {\bibinfo {volume} {22}},\ \bibinfo {pages} {450} (\bibinfo {year}
  {2020})}\BibitemShut {NoStop}%
\bibitem [{\citenamefont {Erlandsson}, \citenamefont {Br{\'o}dka},\ and\
  \citenamefont {Borg}(2017)}]{erlandsson2017seed}%
  \BibitemOpen
  \bibfield  {author} {\bibinfo {author} {\bibfnamefont {F.}~\bibnamefont
  {Erlandsson}}, \bibinfo {author} {\bibfnamefont {P.}~\bibnamefont
  {Br{\'o}dka}}, \ and\ \bibinfo {author} {\bibfnamefont {A.}~\bibnamefont
  {Borg}},\ }\bibfield  {title} {\enquote {\bibinfo {title} {Seed selection for
  information cascade in multilayer networks},}\ }in\ \href@noop {} {\emph
  {\bibinfo {booktitle} {International Conference on Complex Networks and their
  Applications}}}\ (\bibinfo {organization} {Springer},\ \bibinfo {year}
  {2017})\ pp.\ \bibinfo {pages} {426--436}\BibitemShut {NoStop}%
\bibitem [{\citenamefont {Magnani}\ and\ \citenamefont
  {Rossi}(2011)}]{magnani2011ml}%
  \BibitemOpen
  \bibfield  {author} {\bibinfo {author} {\bibfnamefont {M.}~\bibnamefont
  {Magnani}}\ and\ \bibinfo {author} {\bibfnamefont {L.}~\bibnamefont
  {Rossi}},\ }\bibfield  {title} {\enquote {\bibinfo {title} {The ml-model for
  multi-layer social networks},}\ }in\ \href@noop {} {\emph {\bibinfo
  {booktitle} {2011 International Conference on Advances in Social Networks
  Analysis and Mining}}}\ (\bibinfo {organization} {IEEE},\ \bibinfo {year}
  {2011})\ pp.\ \bibinfo {pages} {5--12}\BibitemShut {NoStop}%
\bibitem [{\citenamefont {Br{\'o}dka}\ and\ \citenamefont
  {Kazienko}(2018)}]{Brodka2018}%
  \BibitemOpen
  \bibfield  {author} {\bibinfo {author} {\bibfnamefont {P.}~\bibnamefont
  {Br{\'o}dka}}\ and\ \bibinfo {author} {\bibfnamefont {P.}~\bibnamefont
  {Kazienko}},\ }\enquote {\bibinfo {title} {Multilayer social networks},}\ in\
  \href {\doibase 10.1007/978-1-4939-7131-2_239} {\emph {\bibinfo {booktitle}
  {Encyclopedia of Social Network Analysis and Mining}}},\ \bibinfo {editor}
  {edited by\ \bibinfo {editor} {\bibfnamefont {R.}~\bibnamefont {Alhajj}}\
  and\ \bibinfo {editor} {\bibfnamefont {J.}~\bibnamefont {Rokne}}}\ (\bibinfo
  {publisher} {Springer New York},\ \bibinfo {address} {New York, NY},\
  \bibinfo {year} {2018})\ pp.\ \bibinfo {pages} {1408--1422}\BibitemShut
  {NoStop}%
\bibitem [{\citenamefont {Goldenberg}, \citenamefont {Libai},\ and\
  \citenamefont {Muller}(2001)}]{goldenberg2001talk}%
  \BibitemOpen
  \bibfield  {author} {\bibinfo {author} {\bibfnamefont {J.}~\bibnamefont
  {Goldenberg}}, \bibinfo {author} {\bibfnamefont {B.}~\bibnamefont {Libai}}, \
  and\ \bibinfo {author} {\bibfnamefont {E.}~\bibnamefont {Muller}},\
  }\bibfield  {title} {\enquote {\bibinfo {title} {Talk of the network: A
  complex systems look at the underlying process of word-of-mouth},}\
  }\href@noop {} {\bibfield  {journal} {\bibinfo  {journal} {Marketing
  letters}\ }\textbf {\bibinfo {volume} {12}},\ \bibinfo {pages} {211--223}
  (\bibinfo {year} {2001})}\BibitemShut {NoStop}%
\bibitem [{\citenamefont {Berlingerio}\ \emph {et~al.}(2011)\citenamefont
  {Berlingerio}, \citenamefont {Coscia}, \citenamefont {Giannotti},
  \citenamefont {Monreale},\ and\ \citenamefont
  {Pedreschi}}]{berlingerio2011foundations}%
  \BibitemOpen
  \bibfield  {author} {\bibinfo {author} {\bibfnamefont {M.}~\bibnamefont
  {Berlingerio}}, \bibinfo {author} {\bibfnamefont {M.}~\bibnamefont {Coscia}},
  \bibinfo {author} {\bibfnamefont {F.}~\bibnamefont {Giannotti}}, \bibinfo
  {author} {\bibfnamefont {A.}~\bibnamefont {Monreale}}, \ and\ \bibinfo
  {author} {\bibfnamefont {D.}~\bibnamefont {Pedreschi}},\ }\bibfield  {title}
  {\enquote {\bibinfo {title} {Foundations of multidimensional network
  analysis},}\ }in\ \href@noop {} {\emph {\bibinfo {booktitle} {2011
  international conference on advances in social networks analysis and
  mining}}}\ (\bibinfo {organization} {IEEE},\ \bibinfo {year} {2011})\ pp.\
  \bibinfo {pages} {485--489}\BibitemShut {NoStop}%
\bibitem [{\citenamefont {Salehi}\ \emph {et~al.}(2015)\citenamefont {Salehi},
  \citenamefont {Sharma}, \citenamefont {Marzolla}, \citenamefont {Magnani},
  \citenamefont {Siyari},\ and\ \citenamefont {Montesi}}]{salehi2015spreading}%
  \BibitemOpen
  \bibfield  {author} {\bibinfo {author} {\bibfnamefont {M.}~\bibnamefont
  {Salehi}}, \bibinfo {author} {\bibfnamefont {R.}~\bibnamefont {Sharma}},
  \bibinfo {author} {\bibfnamefont {M.}~\bibnamefont {Marzolla}}, \bibinfo
  {author} {\bibfnamefont {M.}~\bibnamefont {Magnani}}, \bibinfo {author}
  {\bibfnamefont {P.}~\bibnamefont {Siyari}}, \ and\ \bibinfo {author}
  {\bibfnamefont {D.}~\bibnamefont {Montesi}},\ }\bibfield  {title} {\enquote
  {\bibinfo {title} {Spreading processes in multilayer networks},}\ }\href@noop
  {} {\bibfield  {journal} {\bibinfo  {journal} {IEEE Transactions on Network
  Science and Engineering}\ }\textbf {\bibinfo {volume} {2}},\ \bibinfo {pages}
  {65--83} (\bibinfo {year} {2015})}\BibitemShut {NoStop}%
\bibitem [{\citenamefont {Musia\l{}}, \citenamefont {Kazienko},\ and\
  \citenamefont {Br\'{o}dka}(2009)}]{musial2009user}%
  \BibitemOpen
  \bibfield  {author} {\bibinfo {author} {\bibfnamefont {K.}~\bibnamefont
  {Musia\l{}}}, \bibinfo {author} {\bibfnamefont {P.}~\bibnamefont {Kazienko}},
  \ and\ \bibinfo {author} {\bibfnamefont {P.}~\bibnamefont {Br\'{o}dka}},\
  }\bibfield  {title} {\enquote {\bibinfo {title} {User position measures in
  social networks},}\ }in\ \href {\doibase 10.1145/1731011.1731017} {\emph
  {\bibinfo {booktitle} {Proceedings of the 3rd Workshop on Social Network
  Mining and Analysis}}},\ \bibinfo {series and number} {SNA-KDD ’09}\
  (\bibinfo  {publisher} {Association for Computing Machinery},\ \bibinfo
  {address} {New York, NY, USA},\ \bibinfo {year} {2009})\BibitemShut {NoStop}%
\bibitem [{\citenamefont {Hu}\ \emph {et~al.}(2013)\citenamefont {Hu},
  \citenamefont {Gao}, \citenamefont {Ma}, \citenamefont {Yin}, \citenamefont
  {Zhang},\ and\ \citenamefont {Xing}}]{hu2013new}%
  \BibitemOpen
  \bibfield  {author} {\bibinfo {author} {\bibfnamefont {Q.}~\bibnamefont
  {Hu}}, \bibinfo {author} {\bibfnamefont {Y.}~\bibnamefont {Gao}}, \bibinfo
  {author} {\bibfnamefont {P.}~\bibnamefont {Ma}}, \bibinfo {author}
  {\bibfnamefont {Y.}~\bibnamefont {Yin}}, \bibinfo {author} {\bibfnamefont
  {Y.}~\bibnamefont {Zhang}}, \ and\ \bibinfo {author} {\bibfnamefont
  {C.}~\bibnamefont {Xing}},\ }\bibfield  {title} {\enquote {\bibinfo {title}
  {A new approach to identify influential spreaders in complex networks},}\
  }in\ \href@noop {} {\emph {\bibinfo {booktitle} {International Conference on
  Web-Age Information Management}}}\ (\bibinfo {organization} {Springer},\
  \bibinfo {year} {2013})\ pp.\ \bibinfo {pages} {99--104}\BibitemShut
  {NoStop}%
\bibitem [{\citenamefont {Page}\ \emph {et~al.}(1999)\citenamefont {Page},
  \citenamefont {Brin}, \citenamefont {Motwani},\ and\ \citenamefont
  {Winograd}}]{page1999pagerank}%
  \BibitemOpen
  \bibfield  {author} {\bibinfo {author} {\bibfnamefont {L.}~\bibnamefont
  {Page}}, \bibinfo {author} {\bibfnamefont {S.}~\bibnamefont {Brin}}, \bibinfo
  {author} {\bibfnamefont {R.}~\bibnamefont {Motwani}}, \ and\ \bibinfo
  {author} {\bibfnamefont {T.}~\bibnamefont {Winograd}},\ }\href@noop {}
  {\enquote {\bibinfo {title} {The pagerank citation ranking: Bringing order to
  the web.}}\ }\bibinfo {type} {Tech. Rep.}\ (\bibinfo  {institution} {Stanford
  InfoLab},\ \bibinfo {year} {1999})\BibitemShut {NoStop}%
\bibitem [{\citenamefont {Zhang}\ \emph {et~al.}(2016)\citenamefont {Zhang},
  \citenamefont {Chen}, \citenamefont {Dong},\ and\ \citenamefont
  {Zhao}}]{zhang2016identifying}%
  \BibitemOpen
  \bibfield  {author} {\bibinfo {author} {\bibfnamefont {J.-X.}\ \bibnamefont
  {Zhang}}, \bibinfo {author} {\bibfnamefont {D.-B.}\ \bibnamefont {Chen}},
  \bibinfo {author} {\bibfnamefont {Q.}~\bibnamefont {Dong}}, \ and\ \bibinfo
  {author} {\bibfnamefont {Z.-D.}\ \bibnamefont {Zhao}},\ }\bibfield  {title}
  {\enquote {\bibinfo {title} {Identifying a set of influential spreaders in
  complex networks},}\ }\href@noop {} {\bibfield  {journal} {\bibinfo
  {journal} {Scientific reports}\ }\textbf {\bibinfo {volume} {6}},\ \bibinfo
  {pages} {27823} (\bibinfo {year} {2016})}\BibitemShut {NoStop}%
\bibitem [{\citenamefont {Seeman}\ and\ \citenamefont
  {Singer}(2013)}]{seeman2013adaptive}%
  \BibitemOpen
  \bibfield  {author} {\bibinfo {author} {\bibfnamefont {L.}~\bibnamefont
  {Seeman}}\ and\ \bibinfo {author} {\bibfnamefont {Y.}~\bibnamefont
  {Singer}},\ }\bibfield  {title} {\enquote {\bibinfo {title} {Adaptive seeding
  in social networks},}\ }in\ \href@noop {} {\emph {\bibinfo {booktitle} {2013
  IEEE 54th Annual Symposium on Foundations of Computer Science}}}\ (\bibinfo
  {organization} {IEEE},\ \bibinfo {year} {2013})\ pp.\ \bibinfo {pages}
  {459--468}\BibitemShut {NoStop}%
\bibitem [{\citenamefont {Sela}\ \emph {et~al.}(2015)\citenamefont {Sela},
  \citenamefont {Ben-Gal}, \citenamefont {Pentland},\ and\ \citenamefont
  {Shmueli}}]{sela2015improving}%
  \BibitemOpen
  \bibfield  {author} {\bibinfo {author} {\bibfnamefont {A.}~\bibnamefont
  {Sela}}, \bibinfo {author} {\bibfnamefont {I.}~\bibnamefont {Ben-Gal}},
  \bibinfo {author} {\bibfnamefont {A.~S.}\ \bibnamefont {Pentland}}, \ and\
  \bibinfo {author} {\bibfnamefont {E.}~\bibnamefont {Shmueli}},\ }\bibfield
  {title} {\enquote {\bibinfo {title} {Improving information spread through a
  scheduled seeding approach},}\ }in\ \href@noop {} {\emph {\bibinfo
  {booktitle} {Proceedings of the 2015 IEEE/ACM International Conference on
  Advances in Social Networks Analysis and Mining 2015}}}\ (\bibinfo {year}
  {2015})\ pp.\ \bibinfo {pages} {629--632}\BibitemShut {NoStop}%
\bibitem [{\citenamefont {Goldenberg}, \citenamefont {Sela},\ and\
  \citenamefont {Shmueli}(2018)}]{goldenberg2018timing}%
  \BibitemOpen
  \bibfield  {author} {\bibinfo {author} {\bibfnamefont {D.}~\bibnamefont
  {Goldenberg}}, \bibinfo {author} {\bibfnamefont {A.}~\bibnamefont {Sela}}, \
  and\ \bibinfo {author} {\bibfnamefont {E.}~\bibnamefont {Shmueli}},\
  }\bibfield  {title} {\enquote {\bibinfo {title} {Timing matters: Influence
  maximization in social networks through scheduled seeding},}\ }\href@noop {}
  {\bibfield  {journal} {\bibinfo  {journal} {IEEE Transactions on
  Computational Social Systems}\ }\textbf {\bibinfo {volume} {5}},\ \bibinfo
  {pages} {621--638} (\bibinfo {year} {2018})}\BibitemShut {NoStop}%
\bibitem [{\citenamefont {Sela}\ \emph {et~al.}(2018)\citenamefont {Sela},
  \citenamefont {Goldenberg}, \citenamefont {Ben-Gal},\ and\ \citenamefont
  {Shmueli}}]{sela2018active}%
  \BibitemOpen
  \bibfield  {author} {\bibinfo {author} {\bibfnamefont {A.}~\bibnamefont
  {Sela}}, \bibinfo {author} {\bibfnamefont {D.}~\bibnamefont {Goldenberg}},
  \bibinfo {author} {\bibfnamefont {I.}~\bibnamefont {Ben-Gal}}, \ and\
  \bibinfo {author} {\bibfnamefont {E.}~\bibnamefont {Shmueli}},\ }\bibfield
  {title} {\enquote {\bibinfo {title} {Active viral marketing: Incorporating
  continuous active seeding efforts into the diffusion model},}\ }\href@noop {}
  {\bibfield  {journal} {\bibinfo  {journal} {Expert Systems with
  Applications}\ }\textbf {\bibinfo {volume} {107}},\ \bibinfo {pages} {45--60}
  (\bibinfo {year} {2018})}\BibitemShut {NoStop}%
\bibitem [{\citenamefont {He}, \citenamefont {Fu},\ and\ \citenamefont
  {Chen}(2015)}]{he2015novel}%
  \BibitemOpen
  \bibfield  {author} {\bibinfo {author} {\bibfnamefont {J.-L.}\ \bibnamefont
  {He}}, \bibinfo {author} {\bibfnamefont {Y.}~\bibnamefont {Fu}}, \ and\
  \bibinfo {author} {\bibfnamefont {D.-B.}\ \bibnamefont {Chen}},\ }\bibfield
  {title} {\enquote {\bibinfo {title} {A novel top-k strategy for influence
  maximization in complex networks with community structure},}\ }\href@noop {}
  {\bibfield  {journal} {\bibinfo  {journal} {PloS one}\ }\textbf {\bibinfo
  {volume} {10}},\ \bibinfo {pages} {e0145283} (\bibinfo {year}
  {2015})}\BibitemShut {NoStop}%
\bibitem [{\citenamefont {Ni}, \citenamefont {Yang},\ and\ \citenamefont
  {Kong}(2019)}]{ni2019sequential}%
  \BibitemOpen
  \bibfield  {author} {\bibinfo {author} {\bibfnamefont {C.}~\bibnamefont
  {Ni}}, \bibinfo {author} {\bibfnamefont {J.}~\bibnamefont {Yang}}, \ and\
  \bibinfo {author} {\bibfnamefont {D.}~\bibnamefont {Kong}},\ }\bibfield
  {title} {\enquote {\bibinfo {title} {Sequential seeding strategy for social
  influence diffusion with improved entropy-based centrality},}\ }\href@noop {}
  {\bibfield  {journal} {\bibinfo  {journal} {Physica A: Statistical Mechanics
  and its Applications}\ ,\ \bibinfo {pages} {123659}} (\bibinfo {year}
  {2019})}\BibitemShut {NoStop}%
\bibitem [{\citenamefont {Liu}\ and\ \citenamefont
  {Hong}(2018)}]{liu2018sequential}%
  \BibitemOpen
  \bibfield  {author} {\bibinfo {author} {\bibfnamefont {Q.}~\bibnamefont
  {Liu}}\ and\ \bibinfo {author} {\bibfnamefont {T.}~\bibnamefont {Hong}},\
  }\bibfield  {title} {\enquote {\bibinfo {title} {Sequential seeding for
  spreading in complex networks: Influence of the network topology},}\
  }\href@noop {} {\bibfield  {journal} {\bibinfo  {journal} {Physica A:
  Statistical Mechanics and its Applications}\ }\textbf {\bibinfo {volume}
  {508}},\ \bibinfo {pages} {10--17} (\bibinfo {year} {2018})}\BibitemShut
  {NoStop}%
\bibitem [{\citenamefont {Jankowski}\ \emph
  {et~al.}(2017{\natexlab{b}})\citenamefont {Jankowski}, \citenamefont
  {Br{\'o}dka}, \citenamefont {Michalski},\ and\ \citenamefont
  {Kazienko}}]{jankowski2017seeds}%
  \BibitemOpen
  \bibfield  {author} {\bibinfo {author} {\bibfnamefont {J.}~\bibnamefont
  {Jankowski}}, \bibinfo {author} {\bibfnamefont {P.}~\bibnamefont
  {Br{\'o}dka}}, \bibinfo {author} {\bibfnamefont {R.}~\bibnamefont
  {Michalski}}, \ and\ \bibinfo {author} {\bibfnamefont {P.}~\bibnamefont
  {Kazienko}},\ }\bibfield  {title} {\enquote {\bibinfo {title} {Seeds
  buffering for information spreading processes},}\ }in\ \href@noop {} {\emph
  {\bibinfo {booktitle} {International Conference on Social Informatics}}}\
  (\bibinfo {organization} {Springer},\ \bibinfo {year} {2017})\ pp.\ \bibinfo
  {pages} {628--641}\BibitemShut {NoStop}%
\bibitem [{\citenamefont {Jankowski}\ \emph
  {et~al.}(2018{\natexlab{b}})\citenamefont {Jankowski}, \citenamefont
  {Waniek}, \citenamefont {Alshamsi}, \citenamefont {Br{\'o}dka},\ and\
  \citenamefont {Michalski}}]{jankowski2018strategic}%
  \BibitemOpen
  \bibfield  {author} {\bibinfo {author} {\bibfnamefont {J.}~\bibnamefont
  {Jankowski}}, \bibinfo {author} {\bibfnamefont {M.}~\bibnamefont {Waniek}},
  \bibinfo {author} {\bibfnamefont {A.}~\bibnamefont {Alshamsi}}, \bibinfo
  {author} {\bibfnamefont {P.}~\bibnamefont {Br{\'o}dka}}, \ and\ \bibinfo
  {author} {\bibfnamefont {R.}~\bibnamefont {Michalski}},\ }\bibfield  {title}
  {\enquote {\bibinfo {title} {Strategic distribution of seeds to support
  diffusion in complex networks},}\ }\href@noop {} {\bibfield  {journal}
  {\bibinfo  {journal} {PloS one}\ }\textbf {\bibinfo {volume} {13}},\ \bibinfo
  {pages} {e0205130} (\bibinfo {year} {2018}{\natexlab{b}})}\BibitemShut
  {NoStop}%
\bibitem [{\citenamefont {Michalski}, \citenamefont {Jankowski},\ and\
  \citenamefont {Br{\'o}dka}(2020)}]{michalski2020effective}%
  \BibitemOpen
  \bibfield  {author} {\bibinfo {author} {\bibfnamefont {R.}~\bibnamefont
  {Michalski}}, \bibinfo {author} {\bibfnamefont {J.}~\bibnamefont
  {Jankowski}}, \ and\ \bibinfo {author} {\bibfnamefont {P.}~\bibnamefont
  {Br{\'o}dka}},\ }\bibfield  {title} {\enquote {\bibinfo {title} {Effective
  influence spreading in temporal networks with sequential seeding},}\
  }\href@noop {} {\bibfield  {journal} {\bibinfo  {journal} {IEEE Access}\
  }\textbf {\bibinfo {volume} {8}},\ \bibinfo {pages} {151208--151218}
  (\bibinfo {year} {2020})}\BibitemShut {NoStop}%
\bibitem [{\citenamefont {Rossi}\ and\ \citenamefont
  {Magnani}(2015)}]{rossi2015towards}%
  \BibitemOpen
  \bibfield  {author} {\bibinfo {author} {\bibfnamefont {L.}~\bibnamefont
  {Rossi}}\ and\ \bibinfo {author} {\bibfnamefont {M.}~\bibnamefont
  {Magnani}},\ }\bibfield  {title} {\enquote {\bibinfo {title} {Towards
  effective visual analytics on multiplex and multilayer networks},}\
  }\href@noop {} {\bibfield  {journal} {\bibinfo  {journal} {Chaos, Solitons \&
  Fractals}\ }\textbf {\bibinfo {volume} {72}},\ \bibinfo {pages} {68--76}
  (\bibinfo {year} {2015})}\BibitemShut {NoStop}%
\bibitem [{\citenamefont {Coleman}, \citenamefont {Katz},\ and\ \citenamefont
  {Menzel}(1957)}]{coleman1957diffusion}%
  \BibitemOpen
  \bibfield  {author} {\bibinfo {author} {\bibfnamefont {J.}~\bibnamefont
  {Coleman}}, \bibinfo {author} {\bibfnamefont {E.}~\bibnamefont {Katz}}, \
  and\ \bibinfo {author} {\bibfnamefont {H.}~\bibnamefont {Menzel}},\
  }\bibfield  {title} {\enquote {\bibinfo {title} {The diffusion of an
  innovation among physicians},}\ }\href@noop {} {\bibfield  {journal}
  {\bibinfo  {journal} {Sociometry}\ }\textbf {\bibinfo {volume} {20}},\
  \bibinfo {pages} {253--270} (\bibinfo {year} {1957})}\BibitemShut {NoStop}%
\bibitem [{\citenamefont {Cardillo}\ \emph {et~al.}(2013)\citenamefont
  {Cardillo}, \citenamefont {G{\'o}mez-Gardenes}, \citenamefont {Zanin},
  \citenamefont {Romance}, \citenamefont {Papo}, \citenamefont {Del~Pozo},\
  and\ \citenamefont {Boccaletti}}]{cardillo2013emergence}%
  \BibitemOpen
  \bibfield  {author} {\bibinfo {author} {\bibfnamefont {A.}~\bibnamefont
  {Cardillo}}, \bibinfo {author} {\bibfnamefont {J.}~\bibnamefont
  {G{\'o}mez-Gardenes}}, \bibinfo {author} {\bibfnamefont {M.}~\bibnamefont
  {Zanin}}, \bibinfo {author} {\bibfnamefont {M.}~\bibnamefont {Romance}},
  \bibinfo {author} {\bibfnamefont {D.}~\bibnamefont {Papo}}, \bibinfo {author}
  {\bibfnamefont {F.}~\bibnamefont {Del~Pozo}}, \ and\ \bibinfo {author}
  {\bibfnamefont {S.}~\bibnamefont {Boccaletti}},\ }\bibfield  {title}
  {\enquote {\bibinfo {title} {Emergence of network features from
  multiplexity},}\ }\href@noop {} {\bibfield  {journal} {\bibinfo  {journal}
  {Scientific reports}\ }\textbf {\bibinfo {volume} {3}},\ \bibinfo {pages}
  {1344} (\bibinfo {year} {2013})}\BibitemShut {NoStop}%
\bibitem [{\citenamefont {Snijders}\ \emph {et~al.}(2006)\citenamefont
  {Snijders}, \citenamefont {Pattison}, \citenamefont {Robins},\ and\
  \citenamefont {Handcock}}]{snijders2006new}%
  \BibitemOpen
  \bibfield  {author} {\bibinfo {author} {\bibfnamefont {T.~A.}\ \bibnamefont
  {Snijders}}, \bibinfo {author} {\bibfnamefont {P.~E.}\ \bibnamefont
  {Pattison}}, \bibinfo {author} {\bibfnamefont {G.~L.}\ \bibnamefont
  {Robins}}, \ and\ \bibinfo {author} {\bibfnamefont {M.~S.}\ \bibnamefont
  {Handcock}},\ }\bibfield  {title} {\enquote {\bibinfo {title} {New
  specifications for exponential random graph models},}\ }\href@noop {}
  {\bibfield  {journal} {\bibinfo  {journal} {Sociological methodology}\
  }\textbf {\bibinfo {volume} {36}},\ \bibinfo {pages} {99--153} (\bibinfo
  {year} {2006})}\BibitemShut {NoStop}%
\bibitem [{\citenamefont {Magnani}\ and\ \citenamefont
  {Rossi}(2013)}]{magnani2013formation}%
  \BibitemOpen
  \bibfield  {author} {\bibinfo {author} {\bibfnamefont {M.}~\bibnamefont
  {Magnani}}\ and\ \bibinfo {author} {\bibfnamefont {L.}~\bibnamefont
  {Rossi}},\ }\bibfield  {title} {\enquote {\bibinfo {title} {Formation of
  multiple networks},}\ }in\ \href@noop {} {\emph {\bibinfo {booktitle}
  {International Conference on Social Computing, Behavioral-Cultural Modeling,
  and Prediction}}}\ (\bibinfo {organization} {Springer},\ \bibinfo {year}
  {2013})\ pp.\ \bibinfo {pages} {257--264}\BibitemShut {NoStop}%
\bibitem [{\citenamefont {Matteo}, \citenamefont {Davide},\ and\ \citenamefont
  {Mikael}(2020)}]{multinet}%
  \BibitemOpen
  \bibfield  {author} {\bibinfo {author} {\bibfnamefont {M.}~\bibnamefont
  {Matteo}}, \bibinfo {author} {\bibfnamefont {V.}~\bibnamefont {Davide}}, \
  and\ \bibinfo {author} {\bibfnamefont {D.}~\bibnamefont {Mikael}},\ }\href
  {https://cran.r-project.org/web/packages/multinet/index.html} {\emph
  {\bibinfo {title} {Multinet: Analysis and Mining of Multilayer Social
  Networks}}} (\bibinfo {year} {2020})\BibitemShut {NoStop}%
\bibitem [{\citenamefont {Hassan}\ and\ \citenamefont
  {Rashid}(2020)}]{hassan2020operational}%
  \BibitemOpen
  \bibfield  {author} {\bibinfo {author} {\bibfnamefont {B.~A.}\ \bibnamefont
  {Hassan}}\ and\ \bibinfo {author} {\bibfnamefont {T.~A.}\ \bibnamefont
  {Rashid}},\ }\bibfield  {title} {\enquote {\bibinfo {title} {Operational
  framework for recent advances in backtracking search optimisation algorithm:
  A systematic review and performance evaluation},}\ }\href@noop {} {\bibfield
  {journal} {\bibinfo  {journal} {Applied Mathematics and Computation}\
  }\textbf {\bibinfo {volume} {370}},\ \bibinfo {pages} {124919} (\bibinfo
  {year} {2020})}\BibitemShut {NoStop}%
\bibitem [{\citenamefont {De~Domenico}\ \emph {et~al.}(2015)\citenamefont
  {De~Domenico}, \citenamefont {Lancichinetti}, \citenamefont {Arenas},\ and\
  \citenamefont {Rosvall}}]{de2015identifying}%
  \BibitemOpen
  \bibfield  {author} {\bibinfo {author} {\bibfnamefont {M.}~\bibnamefont
  {De~Domenico}}, \bibinfo {author} {\bibfnamefont {A.}~\bibnamefont
  {Lancichinetti}}, \bibinfo {author} {\bibfnamefont {A.}~\bibnamefont
  {Arenas}}, \ and\ \bibinfo {author} {\bibfnamefont {M.}~\bibnamefont
  {Rosvall}},\ }\bibfield  {title} {\enquote {\bibinfo {title} {Identifying
  modular flows on multilayer networks reveals highly overlapping organization
  in interconnected systems},}\ }\href@noop {} {\bibfield  {journal} {\bibinfo
  {journal} {Physical Review X}\ }\textbf {\bibinfo {volume} {5}},\ \bibinfo
  {pages} {011027} (\bibinfo {year} {2015})}\BibitemShut {NoStop}%
\bibitem [{\citenamefont {Iyer}\ and\ \citenamefont
  {Adamic}(2018)}]{iyer2018costs}%
  \BibitemOpen
  \bibfield  {author} {\bibinfo {author} {\bibfnamefont {S.}~\bibnamefont
  {Iyer}}\ and\ \bibinfo {author} {\bibfnamefont {L.~A.}\ \bibnamefont
  {Adamic}},\ }\bibfield  {title} {\enquote {\bibinfo {title} {The costs of
  overambitious seeding of social products},}\ }in\ \href@noop {} {\emph
  {\bibinfo {booktitle} {International Conference on Complex Networks and their
  Applications}}}\ (\bibinfo {organization} {Springer},\ \bibinfo {year}
  {2018})\ pp.\ \bibinfo {pages} {273--286}\BibitemShut {NoStop}%
\bibitem [{\citenamefont {Bass}(1969)}]{bass1969new}%
  \BibitemOpen
  \bibfield  {author} {\bibinfo {author} {\bibfnamefont {F.~M.}\ \bibnamefont
  {Bass}},\ }\bibfield  {title} {\enquote {\bibinfo {title} {A new product
  growth for model consumer durables},}\ }\href@noop {} {\bibfield  {journal}
  {\bibinfo  {journal} {Management science}\ }\textbf {\bibinfo {volume}
  {15}},\ \bibinfo {pages} {215--227} (\bibinfo {year} {1969})}\BibitemShut
  {NoStop}%
\bibitem [{\citenamefont {Rand}\ and\ \citenamefont
  {Rust}(2011)}]{rand2011agent}%
  \BibitemOpen
  \bibfield  {author} {\bibinfo {author} {\bibfnamefont {W.}~\bibnamefont
  {Rand}}\ and\ \bibinfo {author} {\bibfnamefont {R.~T.}\ \bibnamefont
  {Rust}},\ }\bibfield  {title} {\enquote {\bibinfo {title} {Agent-based
  modeling in marketing: Guidelines for rigor},}\ }\href@noop {} {\bibfield
  {journal} {\bibinfo  {journal} {International Journal of research in
  Marketing}\ }\textbf {\bibinfo {volume} {28}},\ \bibinfo {pages} {181--193}
  (\bibinfo {year} {2011})}\BibitemShut {NoStop}%
\bibitem [{\citenamefont {Chica}\ and\ \citenamefont
  {Rand}(2017)}]{chica2017building}%
  \BibitemOpen
  \bibfield  {author} {\bibinfo {author} {\bibfnamefont {M.}~\bibnamefont
  {Chica}}\ and\ \bibinfo {author} {\bibfnamefont {W.}~\bibnamefont {Rand}},\
  }\bibfield  {title} {\enquote {\bibinfo {title} {Building agent-based
  decision support systems for word-of-mouth programs: A freemium
  application},}\ }\href@noop {} {\bibfield  {journal} {\bibinfo  {journal}
  {Journal of Marketing Research}\ }\textbf {\bibinfo {volume} {54}},\ \bibinfo
  {pages} {752--767} (\bibinfo {year} {2017})}\BibitemShut {NoStop}%
\bibitem [{\citenamefont {Scat{\`a}}\ \emph {et~al.}(2016)\citenamefont
  {Scat{\`a}}, \citenamefont {Di~Stefano}, \citenamefont {Li{\`o}},\ and\
  \citenamefont {La~Corte}}]{scata2016impact}%
  \BibitemOpen
  \bibfield  {author} {\bibinfo {author} {\bibfnamefont {M.}~\bibnamefont
  {Scat{\`a}}}, \bibinfo {author} {\bibfnamefont {A.}~\bibnamefont
  {Di~Stefano}}, \bibinfo {author} {\bibfnamefont {P.}~\bibnamefont {Li{\`o}}},
  \ and\ \bibinfo {author} {\bibfnamefont {A.}~\bibnamefont {La~Corte}},\
  }\bibfield  {title} {\enquote {\bibinfo {title} {The impact of heterogeneity
  and awareness in modeling epidemic spreading on multiplex networks},}\
  }\href@noop {} {\bibfield  {journal} {\bibinfo  {journal} {Scientific
  reports}\ }\textbf {\bibinfo {volume} {6}},\ \bibinfo {pages} {1--13}
  (\bibinfo {year} {2016})}\BibitemShut {NoStop}%
\bibitem [{\citenamefont {Zang}(2018)}]{zang2018effects}%
  \BibitemOpen
  \bibfield  {author} {\bibinfo {author} {\bibfnamefont {H.}~\bibnamefont
  {Zang}},\ }\bibfield  {title} {\enquote {\bibinfo {title} {The effects of
  global awareness on the spreading of epidemics in multiplex networks},}\
  }\href@noop {} {\bibfield  {journal} {\bibinfo  {journal} {Physica A:
  Statistical Mechanics and its Applications}\ }\textbf {\bibinfo {volume}
  {492}},\ \bibinfo {pages} {1495--1506} (\bibinfo {year} {2018})}\BibitemShut
  {NoStop}%
\bibitem [{\citenamefont {Holley}\ and\ \citenamefont
  {Liggett}(1975)}]{holley1975ergodic}%
  \BibitemOpen
  \bibfield  {author} {\bibinfo {author} {\bibfnamefont {R.~A.}\ \bibnamefont
  {Holley}}\ and\ \bibinfo {author} {\bibfnamefont {T.~M.}\ \bibnamefont
  {Liggett}},\ }\bibfield  {title} {\enquote {\bibinfo {title} {Ergodic
  theorems for weakly interacting infinite systems and the voter model},}\
  }\href@noop {} {\bibfield  {journal} {\bibinfo  {journal} {The annals of
  probability}\ ,\ \bibinfo {pages} {643--663}} (\bibinfo {year}
  {1975})}\BibitemShut {NoStop}%
\bibitem [{\citenamefont {Bródka}, \citenamefont {Jankowski},\ and\
  \citenamefont {Michalski}(2021{\natexlab{a}})}]{brodka2021github}%
  \BibitemOpen
  \bibfield  {author} {\bibinfo {author} {\bibfnamefont {P.}~\bibnamefont
  {Bródka}}, \bibinfo {author} {\bibfnamefont {J.}~\bibnamefont {Jankowski}},
  \ and\ \bibinfo {author} {\bibfnamefont {R.}~\bibnamefont {Michalski}},\
  }\href {\doibase 10.5281/zenodo.4583675} {\enquote {\bibinfo {title}
  {{Sequential Seeding in Multilayer Networks}},}\ }\bibinfo {howpublished}
  {https://github.com/pbrodka/SQ4MLN} (\bibinfo {year}
  {2021}{\natexlab{a}})\BibitemShut {NoStop}%
\bibitem [{\citenamefont {Bródka}, \citenamefont {Jankowski},\ and\
  \citenamefont {Michalski}(2021{\natexlab{b}})}]{brodka2021sequentialCode}%
  \BibitemOpen
  \bibfield  {author} {\bibinfo {author} {\bibfnamefont {P.}~\bibnamefont
  {Bródka}}, \bibinfo {author} {\bibfnamefont {J.}~\bibnamefont {Jankowski}},
  \ and\ \bibinfo {author} {\bibfnamefont {R.}~\bibnamefont {Michalski}},\
  }\href {\doibase 10.24433/CO.4337573.v1} {\enquote {\bibinfo {title}
  {Sequential seeding in multilayer networks},}\ }\bibinfo {howpublished}
  {\url{https://doi.org/10.24433/CO.4337573.v1}} (\bibinfo {year}
  {2021}{\natexlab{b}})\BibitemShut {NoStop}%
\end{thebibliography}%

\end{document}